\documentclass[lettersize,journal]{IEEEtran}
\usepackage{amsmath,amsfonts}
\usepackage{algorithmic}
\usepackage{algorithm}
\usepackage{array}
\usepackage[caption=false,font=normalsize,labelfont=sf,textfont=sf]{subfig}
\usepackage{textcomp}
\usepackage{stfloats}
\usepackage{url}
\usepackage{verbatim}
\usepackage{graphicx}
\usepackage{cite}
\usepackage{hyperref}

\usepackage[dvipsnames,table,xcdraw]{xcolor}
\usepackage{paralist}
\usepackage{amsmath}
\usepackage{stfloats}

\usepackage{array}
\usepackage{footnote}
\usepackage{booktabs}
\makesavenoteenv{tabular}
\definecolor{myBlack}{rgb}{0.0, 0.0, 0.0}

\begin{document}

\title{TraSculptor: Visual Analytics for Enhanced Decision-Making in Road Traffic Planning}

\author{Zikun Deng, Yuanbang Liu, Mingrui Zhu, Da Xiang, Haiyue Yu, \\ Zicheng Su, Qinglong Lu, Tobias Schreck, and Yi Cai
\thanks{The work was supported by the National Key R\&D Program of China (2022YFE0137800), National Natural Science Foundation of China (62402184), Funding by Science and Technology Projects in Guangzhou (2025A04J3928), and Open Project Program of the State Key Laboratory of CAD\&CG (Grant No. A2407), Zhejiang University.}
\thanks{Z. Deng, Y. Liu, M. Zhu, D. Xiang, H. Yu, and Y. Cai are with (1) the School of Software Engineering, South China University of Technology, and (2) Key Laboratory of Big Data and Intelligent Robot (South China University of Technology), Ministry of Education. Email: zkdeng@scut.edu.cn; \{fqiu639, zhumr1001, kleintravel0209, henryyu0830\}@gmail.com;ycai@scut.edu.cn. Yi Cai is the corresponding author.}
\thanks{Z. Su is with the Key Laboratory of Road and Traffic Engineering of the Ministry of Education, Tongji University. Email: suzicheng@tongji.edu.cn.}
\thanks{Q. Lu is with the Department of Civil and Environmental Engineering, National University of
Singapore, Singapore. Email: qinglong.lu@tum.de.}
\thanks{T. Schreck is with the Graz University of Technology. Email: tobias.schreck@cgv.tugraz.at.}
\thanks{Manuscript received April 19, 2021; revised August 16, 2021.}}

\markboth{Journal of \LaTeX\ Class Files,~Vol.~14, No.~8, August~2021}%
{Shell \MakeLowercase{\textit{et al.}}: A Sample Article Using IEEEtran.cls for IEEE Journals}


\maketitle

\begin{abstract}
The design of urban road networks significantly influences traffic conditions, underscoring the importance of informed traffic planning.
Traffic planning experts rely on specialized platforms to simulate traffic systems, assessing the efficacy of the road network across various states of modifications.
Nevertheless, a prevailing issue persists: many existing traffic planning platforms exhibit inefficiencies in flexibly interacting with the road network's structure and attributes and intuitively comparing multiple states during the iterative planning process.
This paper introduces TraSculptor, an interactive planning decision-making system. To develop TraSculptor, we identify and address two challenges: interactive modification of road networks and intuitive comparison of multiple network states.
For the first challenge, we establish flexible interactions to enable experts to easily and directly modify the road network on the map.
For the second challenge, we design a comparison view with a history tree of multiple states and a road-state matrix to facilitate intuitive comparison of road network states.
To evaluate TraSculptor, we provided a usage scenario where the Braess's paradox was showcased, invited experts to perform a case study on the Sioux Falls network, and collected expert feedback through interviews.

\end{abstract}


\section{Introduction}
\IEEEPARstart{I}{nformed} traffic planning is a prerequisite for an efficient transportation system, and any inefficient planning will bring huge economic losses~\cite{gao2024evaluation}.
One of the important traffic planning tasks is to design or improve a road network~\cite{merchan2020quantifying,antunes2003accessibility}, where each node is an intersection, and each directed edge is a road segment from one node to another.
This paper uses ``road'' to refer to ``road segment'' for simplicity.
Each road has an attribute called capacity that determines how many vehicles can freely pass through the road~\cite{meng2008general}.
An optimal road network has the best traffic efficiency given the traffic OD demand~\cite{ccolak2016understanding}.

Traffic planning experts often utilize traffic simulations to come up with and evaluate every potential strategy (e.g., improving the capacity of a road by expanding the road) in a trial-and-error manner~\cite{LIU2020102939,MATSim}, rather than modifying traffic facilities in urban spaces realistically.
For example, the expert first investigates the traffic network to identify the inefficient road based on the observed traffic measurements (e.g., speed and volume), then improves the road's capacity by expanding it, and evaluates whether the whole transportation system becomes more efficient based on the simulated traffic volumes.

Many traffic planning platforms have been developed to assist the experts in designing or improving road networks.
However, the interactions supported still rely largely on basic UI widgets such as clicks and input boxes.
The user cannot interact with the road network and modify it intuitively and conveniently.
Moreover, the road network design process is usually iterative and trial-and-error, i.e., the expert needs to modify the network that has been modified in the previous steps.
The current platforms do not support the user to efficiently perform iterative operations such as tracking, rolling back, and opening branches between various design steps.
Finally, the iterative process will generate multiple states of the road network.
None of the platforms enables the user to compare these states effectively and thereby make informed decisions.
These limitations can be addressed by visual analytics approaches where intuitive visualizations and interactions are combined to empower expert users~\cite{DBLP:series/lncs/KeimAFGKM08,DBLP:journals/vc/LiuCWL14,DBLP:journals/cvm/DengWLTXW23}.
However, no existing approaches focused on road traffic planning scenarios.

We aim to propose a visual analytics approach to assist experts in interactive, iterative, and informed road traffic planning.
Developing such an approach should address the following two challenges:

\textbf{Modifying the road network interactively.}
Real-world countermeasures like road construction, expansion, and closure correspond to attribute- or structure-level modifications of the abstract road network.
Effective modification interactions should not only enable direct manipulation of the network but also align with domain experts’ intrinsic understanding of real-world traffic planning practices, thereby establishing intuitive interaction metaphors~\cite{NEALE1997441, Maa1992}.
However, existing approaches to network interaction primarily focus on adjusting or exploring network layouts~\cite{DBLP:journals/tvcg/WangWSZLFSDC18,DBLP:journals/tvcg/PanCZZZZCFW21} and fail to incorporate traffic planning metaphors~\cite{DBLP:journals/informatics/EichnerGST16}, limiting their applicability in practical traffic planning scenarios.

\textbf{Comparing multiple states intuitively.}
The modification process of a road network is often iterative and multi-branched, generating multiple road networks with different states.
Experts need to trace the states of multiple road networks and compare them.
However, comparing multiple geospatial road networks regarding the associated information of every edge and node is challenging.
It is crude to extract quantitative indicators (e.g., the average speed of the entire road network) to rank these states since the roads that are getting worse may be compromised by the roads that are getting better.
Some existing geo-network visualizations can support structural comparison~\cite{DBLP:journals/tvcg/DengWXBZXCW22,deng2022multilevel,DBLP:journals/tvcg/WangLYZW13} but not the attribute-level comparison of many edges or nodes.

This study presents an interactive planning decision-making system called TraSculptor (the abbreviation of traffic network ``sculptor'') to address the aforementioned challenges.
For the first challenge, taking roads as examples, we worked closely with domain experts and identified four countermeasures, each targeting roads.
We further established a mapping between these countermeasures and structural- or attribute-level modifications in the abstract road network based on traffic planning metaphors.
Finally, a set of intuitive interactions is developed and included in TraSculptor, enabling the user to modify the road network on the map directly and flexibly.
For the second challenge, we designed a history tree of multiple states to facilitate tracing the iterative modification process and a road-state matrix to organize changes in the road by every state for an intuitive comparison of multiple states.
The history tree and the road-state matrix are aligned by the states and integrated into TraSculptor.
TraSculptor is evaluated using usage scenarios performed by domain experts and positive expert feedback collected from expert interviews.

The contributions can be summarized as follows:
\begin{compactitem}[$\diamond$]
    \item We characterize the domain problem of road traffic planning and compile it into a three-stage decision-making framework with six requirements.
    \item We designed user interactions mapped from road-level real-world countermeasures that enable intuitive and direct simulation of the road network on the map.
    \item We design a novel comparison view that facilitates the comparison among multiple road network states during the iterative planning process;
    \item We develop TraSculptor, to the best of our knowledge, the first visual analytics system for road traffic planning, available at \url{https://github.com/akihaisland/TraSculptor}.
\end{compactitem}

\section{Related Work}
This section discusses two parts of related work: (1) traffic simulation and planning platforms, (2) traffic visualization and visual analytics, and (3) network interaction and comparison.

\subsection{Traffic Simulation and Planning Platforms}
\textbf{Traffic simulation} is effective in evaluating new policies and countermeasures during the traffic planning process.
Based on the spatial scale, traffic simulation can be classified into macroscopic, mesoscopic, and microscopic types~\cite{kotusevski2009review,moller2014introduction}.

\textit{Macroscopic} simulation models focus on the traffic assignment within a city, intending to analyze and replicate the macro-level features of traffic flow.
The key idea is, given a road network and the traffic demand between each pair of traffic nodes, to assign or simulate the vehicles running on each road.
The simulation serves as a means to evaluate the overall performance of the existing and planned road networks~\cite{su2023hierarchical} as well as the public transit system~\cite{chow2021adaptive}.
\textit{Mesoscopic} simulation models shift the attention to vehicle groups as the subjects.
Lighthill and Whitham~\cite{lighthill1955kinematic} proposed one of the classic models
It treats traffic flow as a fluid to simulate vehicle behaviors based on a fluid-dynamical description.
This idea is widely followed.
\textit{Microscopic} simulation takes individual vehicles as the subject.
It simulates driving behaviors and interactions between vehicles based mainly on the car-following model~\cite{su2021adaptive} or lane-changing model~\cite{moridpour2010lane}.

In this study, we focus on the macroscopic simulation, i.e., the traffic assignment that works on traffic road networks.

\textbf{Traffic planning platforms} are important means for domain experts to use traffic simulation models and accomplish traffic planning tasks.
The most popular platforms with macroscopic simulation include TransCAD~\cite{TransCAD}, EMME~\cite{EMME}, PTV-Visum~\cite{PTV-Visum}, and so on.
They all superimpose the traffic visualization of speed or volume on the map to show the simulation results.
Furthermore, they allow users to compare two simulation results over the road network by computing the differences and then using the same visual encodings as one result.
However, they cannot support iterative and continuous modification of the road network, nor can they support the comparison of multiple states of the road network (i.e., multiple simulation results).

Traffic planning needs to be enhanced by user-friendly interactions and intuitive visualization, toward better traffic planning decisions.

\subsection{Traffic Visual Analytics and Visualization~\label{sec:trafficVA}}

\textbf{Traffic visual analytics} combines traffic visualizations, user interactions, and advanced urban computing models to serve human-in-the-loop domain applications, such as traffic congestion tracing~\cite{DBLP:journals/tvcg/DengWLBZSXW22,DBLP:journals/tvcg/DengWXBZXCW22}, mobility analysis~\cite{DBLP:journals/tvcg/LandesbergerBRA16}, route planning~\cite{DBLP:journals/tvcg/WengZDMBZXW21,DBLP:journals/tvcg/LorenzoSCBPN16,DBLP:journals/jvis/DengWW23,DBLP:conf/visualization/LiuLTLMC20}, traffic prediction~\cite{DBLP:journals/tvcg/JinLPCTCK23}, and location selection~\cite{FSLens,DBLP:journals/tvcg/WengCDWCW19}.
The traffic planning we are studying is a decision-making problem in the transportation domain.
The three stages (intelligence, design, and choice)~\cite{simon1960new} for informed decision-making can be well-supported by visual analytics~\cite{DecisionMaking,DBLP:journals/ivs/HanS23}.
Traffic visual analytics has many applications in decision-making.
In AllAborad~\cite{DBLP:journals/tvcg/LorenzoSCBPN16}, an algorithm extends/expands the existing transit network based on the OD demands, and then interactive visualization helps experts compare the new network with the original one but mainly relies on numerical metrics.
ShuttleVis~\cite{DBLP:conf/visualization/LiuLTLMC20} facilitated experts to compare multiple automatically-generated bus routes from multiple numeric metrics.
Beyond the comparison of candidates, BNVA~\cite{DBLP:journals/tvcg/WengZDMBZXW21} enabled experts to interactively steer the Monte Carlo tree search and extend the bus routes.
cT$^3$~\cite{DBLP:journals/jvis/DengWW23} supported tourists planning ideal travel routes by interactively cutting, splicing, and combining prior tourists' travel routes.

Existing visual analytics did not pay attention to road traffic planning and were difficult to apply in this scenario.
Current interaction methods are constrained to route or line-based forms, which focus on selecting~\cite{DBLP:conf/visualization/LiuLTLMC20,DBLP:journals/tvcg/WengZDMBZXW21} or organizing~\cite{DBLP:journals/jvis/DengWW23} sequences of locations.
In our context, interactions should be network-oriented, allowing users to modify road networks—for example, by adding a new edge to simulate road construction.
Besides, existing comparative visualizations, which provide statistical summaries of individual edges, are insufficient for detailed road network comparisons. They lack the capability to examine edge-level attributes in the depth required for effective traffic planning.

\textbf{Traffic visualizations} aim to visually present the traffic data for understanding, sense-making, and reasoning~\cite{TrafficSurvey,DBLP:journals/cvm/DengWLTXW23}.
Non-spatial traffic data can be visualized in many forms,  not limited to the geographic map.
Take the public traffic system as an example: passenger numbers can be visualized as OD matrix~\cite{DBLP:journals/tvcg/WengZDMBZXW21},
and the schedule delay can be exposed via Marey graph~\cite{BusSchedules}, and the efficiency can be analyzed with isotime flow map for the efficiency~\cite{ZengPublicTraffic}.
Spatial traffic data are usually presented via map-based visualization for better understanding within the spatial context.
Sun et al.~\cite{DBLP:journals/tvcg/SunLQW17} allowed temporal traffic data visualizations to be embedded into the road network.
Lee et al.~\cite{CongestionVis} superimposed the traffic information (e.g., speed) on the road network on the map as most navigation platforms do, but cannot support comparison among multiple such networks.
For comparing multiple road networks at the structural level, Wang et al.~\cite{DBLP:journals/tvcg/WangLYZW13} adopted mini maps, while Deng et al.~\cite{deng2022multilevel} stacked OD trips on the map by their shared edges.
Nonetheless, comparing multiple road networks at the attribute levels remains unexplored given existing geo-network visualizations~\cite{DBLP:journals/cgf/SchottlerYPB21,DBLP:journals/tvcg/DengWXBZXCW22}.

In this study, we follow the approach in most navigation platforms to superimpose the visualization of traffic data (e.g., traffic speed and volume) onto the road network according to the road.
The challenging problem is comparing the attributes of multiple such networks.

\subsection{Network Interaction and Comparison}
As mentioned above, iterative road traffic planning involves interacting with and comparing road networks.

\textbf{Interacting} with a road network in our scenario involves modifying the network to create candidates.
Platforms in the field offer standard user interfaces (UIs) for these modifications, typically allowing users to click on UI elements or input commands with the keyboard.
Current methods in the HCI and visualization communities for interacting with networks primarily focus on adjusting or exploring their visual layout~\cite{DBLP:journals/tvcg/WangWSZLFSDC18,DBLP:journals/tvcg/PanCZZZZCFW21}, rather than enabling detailed modifications of individual components.
An exception, Eichner et al.'s method~\cite{DBLP:journals/informatics/EichnerGST16} allows users to modify both the structure and attributes of networks, making it the most relevant to our scenario.
However, it fails to leverage traffic planning metaphors.
Research has shown that interactions inspired by real-world metaphors—aligning closely with traffic planning practices—are generally more intuitive and effective~\cite{NEALE1997441, Maa1992}.
For example, building a new road should be like drawing it on a map.

In our study, we collaborated with domain experts to design intuitive interaction methods by incorporating real-world traffic planning metaphors into network modification tasks.

\textbf{Comparing} road network states obtained by multiple simulations is a prerequisite for wise decision-making.
The first step is to visually trace the iterative planning process.
Many analytical provenance visualizations~\cite{DBLP:conf/iv/MeninCFCW21,DBLP:journals/cgf/XuOWSCW20,DBLP:journals/tvcg/RaganESC16,DBLP:journals/tvcg/Nguyen0WWAF16} focused on this type of task. 
Most of them present the history of states with glyphs either along the timeline~\cite{DBLP:journals/tvcg/Nguyen0WWAF16,DBLP:conf/uist/GrossmanMF10} or within a node-link diagram~\cite{DBLP:journals/tvcg/BorkinYBMGSP13,DBLP:journals/tvcg/StitzGPZS19}.
We follow these ideas to clearly present the relationship between historical states.
However, road networks with geographic context, structural information, and associated attributes cannot be effectively abstracted into a single glyph for effective understanding and comparison.

Effective comparative visualizations have been developed for network comparison.
For general networks, approaches such as node aggregation~\cite{DBLP:journals/tvcg/YangSDTLT17} and specialized network layouts~\cite{DBLP:journals/jvis/JinC0QXC21} facilitate pairwise comparisons.
In scenarios involving multiple networks, small multiples with side-by-side comparison capability are an effective strategy ~\cite{DBLP:journals/tvcg/ShiHTTDJWT22,DBLP:journals/tvcg/YoghourdjianDKM18}.
Dynamic networks, due to their inherent temporal nature, are often compared using visualizations closely integrated with timelines~\cite{DBLP:conf/vinci/BurchW14,DBLP:journals/cgf/BachRDMFG15}.
As for geo-networks, they are either plotted as mini-maps~\cite{DBLP:journals/tvcg/WangLYZW13} or superimposed on the same map in an edge-stacking manner~\cite{DBLP:journals/cgf/CornelKSHBVW16,deng2022multilevel} for comparison.
These visualizations mainly support structural comparison through network transformation and optimized layouts.
They struggle to reveal differences in multiple associated edge attributes, e.g., the traffic volume on the road, across multiple geo-networks, and aligning them with existing provenance visualizations remains challenging.

In our study, we first employ a history tree following the existing node-link diagram for analytical provenance.
We also design a novel matrix-like comparison view with individual road network states spread out vertically in columns and multiple states aligned horizontally with the history tree, enabling attribute-based comparisons of networks within the provenance process.

\section{Overview}
This section formally defines important concepts, introduces cooperation with domain experts, introduces the traffic assignment, and summarizes domain experts' requirements.

\subsection{Concepts~\label{sec:concepts}}
\begin{figure}[tb!]
  \centering 
  \includegraphics[width=\columnwidth]{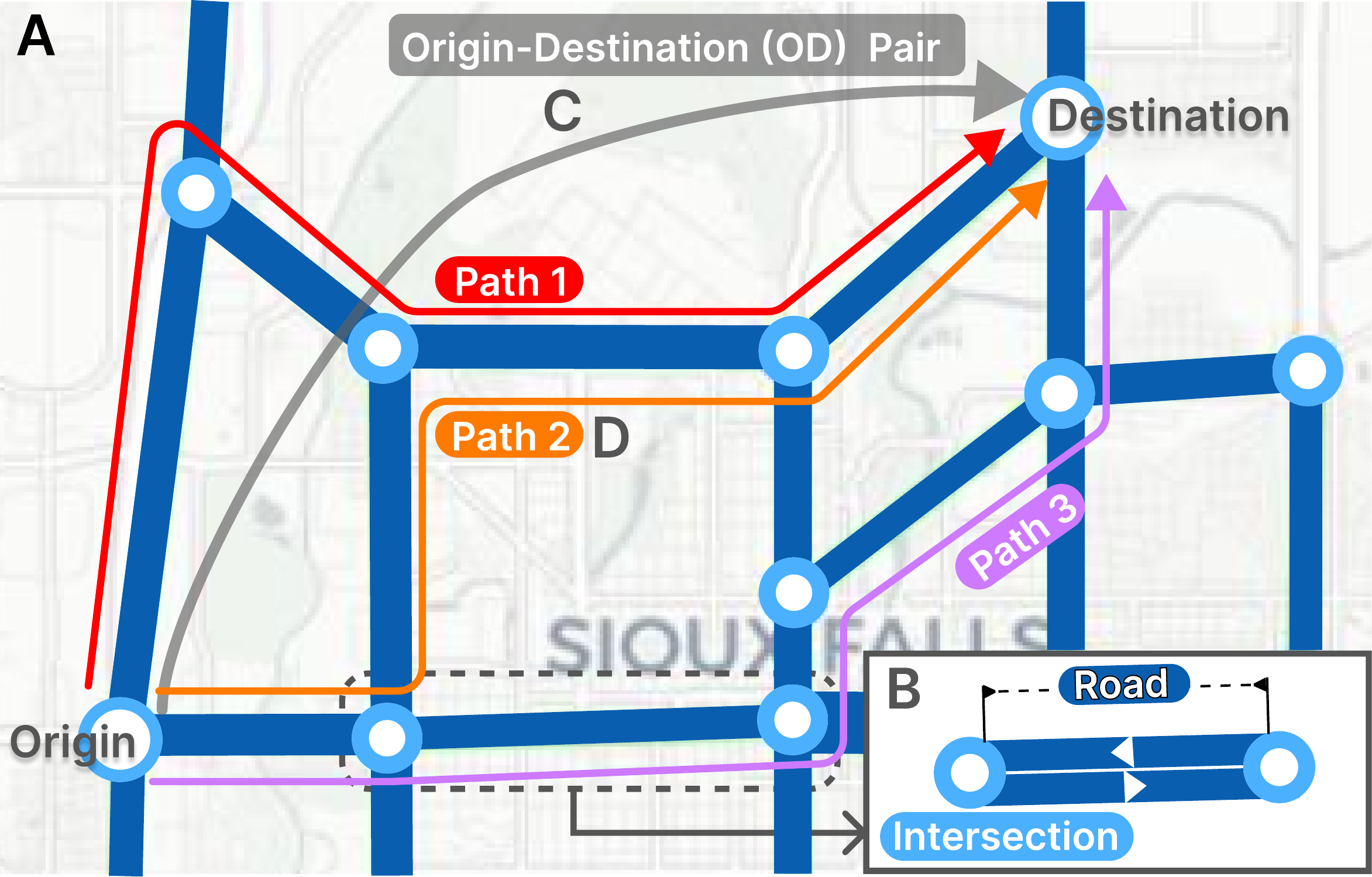}
  \caption{%
  Concept illustration. (A) A road network comprises (B) intersections and roads. In compliance with local traffic regulations, we set the road network in right-hand traffic. (C) An OD pair with an origin and a destination, which is associated with the traffic demand between the origin and destination. (D) One of the paths from the origin to the destination. 
  }
  \label{fig:RN}
\end{figure}

Macro traffic simulation models individuals' travel choices within a road network based on their travel demands, thereby determining traffic conditions through the number of individuals on each road (Section \ref{Sec:model}).
Traffic conditions within a road network will differ before and after (candidate) modifications are made.
The traffic planning process should identify the optimal modification to enhance overall traffic efficiency.
Relevant concepts are introduced as follows:
\vspace{3pt}

    A \textbf{road network} is a direct graph $G = (N, E)$ (\autoref{fig:RN}A).
    Each node $n_i \in N$ is an intersection with a geographic position, and each directed edge $e \in E$ is a road connecting two intersections (\autoref{fig:RN}B).
    Each road has four static attributes: 1) start intersection, 2) end intersection, 3) road capacity $f^e$, and 4) free flow travel time (FFTT) $t_f^e$.
    In particular, the road capacity measures the maximum number of vehicles that may pass through per unit of time.
    Each road also has two dynamic attributes:
    5) actual traffic volume (or actual traffic flow) $f_a^e$ and 6) actual travel time $t_a^e$.
    The road \textbf{status} is determined by these attributes.
    The statuses of all roads reflect the road network \textbf{state}.
    In the traffic planning scenario, each candidate modification on the road network can generate a new state.

    An OD pair $o = (n_r, n_s)$ is a tuple of two intersections $n_r$ and $n_s$. $n_r$ is the origin and $n_s$ the destination (\autoref{fig:RN}C).

    An \textbf{OD trip}, the smallest unit in the traffic assignment, means a person needs to travel between an OD pair.

    The \textbf{travel demand} is the number of OD trips between an OD pair, which is usually obtained via travel surveys~\cite{mcnally2007four} or traffic-measurement-based OD estimation methods~\cite{lu2024two}.

    A \textbf{path} $p = \{\dots, e, \dots\}$ is a set of ordered roads that describes how to move from the origin to the destination of an OD pair (\autoref{fig:RN}D).
    There can be multiple paths \textbf{P}$_o = \{\dots, p^i, \dots\}$ for an OD pair $o$.
    The number of vehicles (actual volume) that pass $p$ is denoted as $f_a^p$.
    The actual travel time for a vehicle to go through the path is denoted as $t_a^p$.
    It is estimated by the sum of the actual travel time of each road belonging to the path $p$.

\vspace{3pt}
The concepts of the road network and travel demands (and OD trips) are further clarified under practical applications as follows.
First, the complete road network is unnecessary.
For instance, low-level roads (e.g., alleyways) are usually excluded.
Similarly, when planning for highways, only the highways themselves are considered due to their closed nature.
Second, travel demands can be achieved via classic travel surveys~\cite{mcnally2007four} or advanced traffic-measurement-based OD estimation methods~\cite{lu2024two}, and the spatial granularity of travel demand varies; it can range from demand between blocks to demand between streets.
For ease of calculation, the spatial positions of travel demands (and OD trips) are typically offset to the nearest intersection~\cite{ wang2013global}.
For this reason, the size of the dataset depends on the size of the road network.

\subsection{Traffic Assignment~\label{Sec:model}}

Traffic assignment is to assign each OD trip to the road network by simulating the driver's choice, finally obtaining the road status in the road network.
Intuitively, every driver tends to choose the path with the shortest travel time.
However, the more vehicles there are, the greater the actual travel time is than the free flow travel time (FFTT).
Based on the widely used BPR function~\cite{spiess1990conical,BRP}, $t_a^e$ depends on $f_a^e$:
\[t_a^e(f_a^e) = t_f^e
\times 
(1 + 0.15 \times (f_a^e / f^e)^4).  \]
The well-established SUE (\textbf{S}tochastic \textbf{U}ser \textbf{E}quilibrium) model~\cite{sheffi1982algorithm} assumes that the driver cannot fully and timely know the actual travel times of each path, and thus, describes the path selection probability with the Logit selection model:
\[
Pro^p = \frac{exp(-\theta t_a^p)}{\sum_{\hat{p} \in \bf{P}_o} exp(-\theta t_a^{\hat{p}})}, p \in \bf{P}_o.
\]
With this equation, $\theta > 0$ can be explained as the degree of the user's discomfort with the travel time~\cite{liu2009method}.
The smaller $\theta$, the more random the assignment, and vice versa.
We follow one of the most classic studies~\cite{sheffi1982algorithm} to set $\theta$ to 0.3. 
The traffic assignment is essentially an optimization problem where every driver in the travel demands chooses (is assigned to) the route they think is fastest.
We adopt the Self-Regulated Averaging method~\cite{sheffi1982algorithm,liu2009method} to solve this optimization problem.
Please refer to the appendix for more information.

In sum, given a road network with static attributes, and a set of OD pairs associated with OD trips, the trips are assigned to the paths and roads of the network, deriving the actual traffic volume $f_a^e$ and actual travel time $t_a^e$ for each road $e$.

\subsection{Cooperation}
This study resulted from the cooperation between transportation engineering and computer science, and follows the nine-stage methodological framework of the design study~\cite{DBLP:journals/tvcg/SedlmairMM12}: 
\textit{Learn, Winnow, Cast, Discover, Design, Implement, Deploy, Reflect, and Write.}
Below, [*] is used to indicate some of the stages in this framework.

Due to the rapid development of data visualization and human-computer interaction techniques, existing traffic analysis platforms (e.g., bus network planning~\cite{DBLP:journals/tvcg/LorenzoSCBPN16,DBLP:journals/tvcg/WengZDMBZXW21} and traffic congestion analysis~\cite{DBLP:journals/tvcg/DengWLBZSXW22,DBLP:journals/tvcg/PiYSJ21}) tend to be human-computer collaborative.
\textbf{[Winnow]} A team of three experts in the field of transportation engineering turned to the visual analytics community for human-machine collaboration solutions for traffic planning.
Two experts (EA and EB) are young professors with more than five years of experience in transportation engineering.
Another expert (EC) is a five-year Ph.D. candidate and focuses on simulation-based traffic system optimization.
We convened meetings many times to interview them, thereby understanding their workflow in traffic planning, as well as the inconveniences they encounter in terms of interaction and analysis.

In the first meeting, we follow Cibulski et al.'s interview guideline~\cite{cibulski2022supporting} to \textbf{[Learn]} understand the expert's routine and tacit workflow for \textbf{[Discover]} requirement characterization.
Every time government departments commission experts to conduct road traffic planning, experts need to use traffic demand data and road networks to perform traffic analysis and simulation.
We asked the experts to share their traffic planning experiences in detail, and we tried to conclude their tacit workflow into a three-stage framework that can be improved via human-computer interactions and visual analytics.
First, they needed to identify traffic inefficiency, such as bottlenecks, which could be derived from traffic data analysis or pointed out by the local traffic department.
Subsequently, they needed to propose candidate solutions for traffic improvement based on TransCAD's simulation capability.
They particularly mentioned that multiple candidates might exist, and they needed to determine the most suitable one and persuade the departments why others were less favorable.

\textbf{[Design]} In subsequent meetings, we confirmed with experts that their tacit workflow essentially follows the well-established three-stage decision-making framework~\cite{simon1960new}.
We designed a prototype visual analytics system according to this framework and presented it to experts to verify whether it aligned with their workflow as well as their cognitive and usage habits, which in turn refined the prototype and clarified requirements.
Traffic countermeasures were also condensed during these meetings and were broken down into several basic modifications in Section \ref{sec:modifications}.
\textbf{[Implement \& Deploy]} The above design process is iterated through the validation of the prototype system with bi-weekly meetings with the domain experts. 
In particular, during this stage, we obtain usage scenarios performed by the experts, which evaluate the proposed solution.
Finally, we share some lessons learned from this design study with researchers of visual analytics.

\subsection{Requirement Analysis~\label{Sec:req}}  
The requirements of experts are summarized following the three-stage decision-making framework: intelligence, design, and choice~\cite{simon1960new}:

    \vspace{3pt}
    \noindent \textbf{INTELLIGENCE.} In this stage, the user analyzes the available data to understand the current state of a traffic system, such as which part is bad and why. Subsequent road network modifications will revolve around this part.

    \textbf{R1: Intuitive presentation of the traffic status.}
    The experts commented that presenting the traffic status is of importance in the whole planning process.
    First, identifying defective roads based on the traffic status, such as roads with very low traffic speeds and excessive traffic volume, is the preliminary step of traffic planning.
    Second, evaluating the planning results on the roads also requires the traffic status to be visually displayed.
    Moreover, these analyses should be performed based on the spatial context.

    \textbf{R2: On-demand investigation of the traffic demands.}
    After identifying defective roads, the experts need to learn where the cars come from that cause the bad status on the roads. Such information will help experts add new roads and route traffic.
    The experts commented it would be better to follow an exploratory analysis that discovers first and explains later because 1) the visualization of massive OD pairs or OD trips is non-trivial, and 2) the existing effective visualizations~\cite{deng2022multilevel,DBLP:journals/tvcg/YangDGM17} may overwhelm users or distract them from the already complicated workflow of traffic planning.

    \vspace{3pt}
    \noindent \textbf{DESIGN.} 
    In this stage, the user issues targeted modifications to the system based on domain knowledge and the results of the previous intelligence stage rather than by enumeration to generate candidates.
    In our study, each candidate is a new road network generated by users when editing the network structure or network attribute.

    \textbf{R3: Interactive edits of the road network.}
    The main goal of traffic planning is to evaluate which countermeasure is best to improve the transportation system.
    Multiple real-world countermeasures (e.g., expanding the road or building a new road) can be naturally mapped to edits of the abstract road network.
    Thus, the experts hope to edit the road network flexibly for various countermeasure simulations.
    They also prefer that the editing interactions match real-world concepts and act on the road network directly on the map.    
    
    \textbf{R4: Iterative process of the road network editing.}
    The planning process is inherently iterative.
    For example, to induce the traffic flow, a new road may need to be added between two intersections, and afterward, an existing two-way road may need to be changed to one-way.
    If the result is unsatisfactory, the user may turn back and choose to add the road between another pair of two intersections.
    Hence, the experts need to browse the history of edits and start new branches of edits from any state of history. 

    \vspace{3pt}
    \noindent \textbf{CHOICE.} In this stage, the user evaluates and compares candidates to select the optimal one, which usually involves multiple criteria.

    \textbf{R5: Comparative analysis of simulation results.}
    To make informed decisions, the experts should evaluate candidate road networks and compare them, considering the costs of countermeasures and travel-related performances.
    These criteria are required to be visually displayed for them.
    Furthermore, since each road network is a geospatial network, it would be better if the comparative analysis could be performed closely with the spatial context.

    \textbf{R6: Easy Identification of Interesting Roads.}
    A road network may contain many roads, but not all of them require comparison.
    Specifically, incremental traffic planning (i.e., upgrading an existing road network) will not affect the whole road network but only the local part of the road network.
    The intuition is that a road is not affected by a road being repaired tens of kilometers away.
    The experts are mainly interested in roads that exhibit obvious changes in key attributes across the states generated in the iterative design process.
    These attributes include actual traffic volume and actual travel time, which are critical indicators of road performance.

\section{Visual Design~\label{Sec:design}}
This section introduces the visual design of TraSculptor that satisfies the aforementioned requirements.

\begin{figure*}[tb!]
  \centering 
  \includegraphics[width=\linewidth]{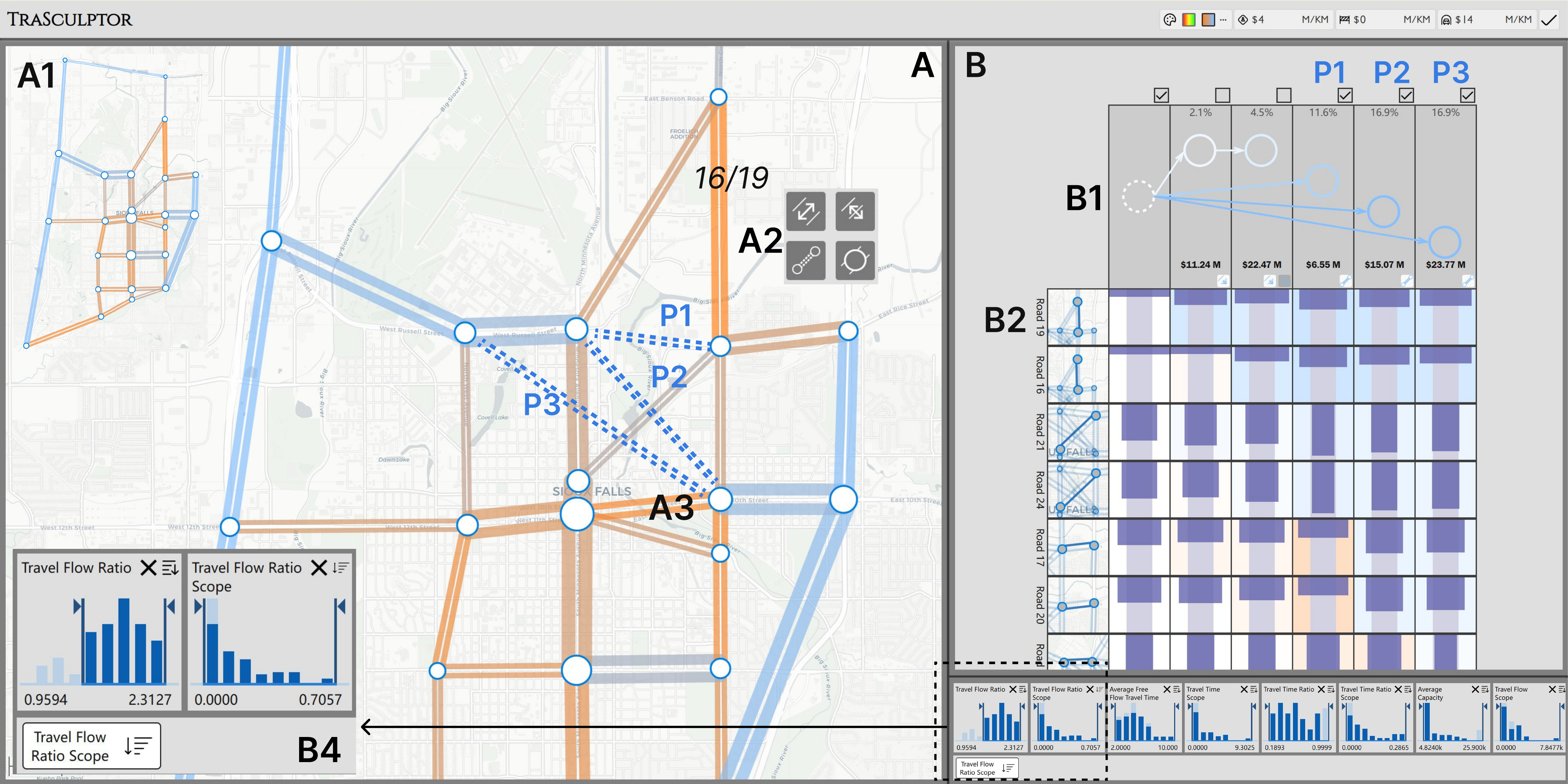}
  \caption{%
  The system interface of TraSculptor. (A) The map view displays the road network with a node-link diagram and visually presents the traffic status of each road. It enables users to directly modify the road network within the spatial context.
    (B) The comparison view comprises (B1) a history tree of the iterative modification process, (B2) a road-state matrix for comparing multiple states of the road network at the road level, and (B3) an indicator panel for filtering and ranking the roads.
    (P1, P2, and P3) Three candidate modifications are indicated by dashed blue lines, each of which adds a new road (P1) or tunnel (P2 and P3) into the road network.
  }
  \label{fig:teaser}
\end{figure*}

\subsection{Overview and Workflow}
TraSculptor is a human-machine collaborative system where users can design and improve the road network with the aid of interactive visualizations and the real-time traffic assignment module.
The system comprises two views: map view and comparison view.

The map view (\autoref{fig:teaser}A) presents the traffic state of the road network and enables interactive road network modification.
Given the initial road network, the user analyzes its state and obtains the inefficient parts in the map view (\textbf{R1}).
The user then interactively modifies the road network in the map view to simulate real-world countermeasures (\textbf{R3}).
A new state is generated based on the modified network and traffic assignment module and is also visualized in the map view.

Since the modification is an iterative process, the system includes the comparison view (\autoref{fig:teaser}B) with a history tree of modifications and a road-state matrix for comparative analysis.
In particular, the history tree (\autoref{fig:teaser}B1) enables users to go through the road network states generated during the iterative modifications (\textbf{R4}).
If the current state is unsatisfactory, the user can roll back and start a new branch of modifications.
The road-state matrix (\autoref{fig:teaser}B2) is aligned with the history tree by the state and supports a side-by-side comparison of multiple roads over states (\textbf{R5}).
The user can decide which state is the best based on the comparative analysis.
Flexible ranking and filters are implemented for users to identify the roads of their interests (\textbf{R6}).

\subsection{Map View}
The map view (\autoref{fig:teaser}A) mainly comprises a road network visualization to present the traffic state of the road network (\textbf{R1}).
Based on the road network, the user can inspect the OD demands by roads to inform the network modifications (\textbf{R2}).
Furthermore, the road network visualization is equipped with flexible and intuitive interactions, enabling the modifications on the road network directly (\textbf{R3}).

\subsubsection{Visualizing the Road Network}
The spatial context of the road network is the most important.
Following most prior studies, we adopt the geographic map to provide the spatial context of the road network.
Each road network can be decomposed into the structure and attributes.

\textbf{Structure.}
In the map view, we use the node-link diagram to visualize the structure of a road network.
Each node is an intersection, and each directed link between the nodes is a road.
After intersections are plotted on the map according to the geographic position of the intersections, the roads are exactly mapped on the map.
There can be two roads between two intersections because the traffic flows are in both directions.
The two roads are placed following local traffic rules (i.e., drive on the right or on the left).

Note that, for simplicity, in the context of system design and usage, ``\textit{road}'' can refer to a graphical object within the node-link diagram in the map view, acting as a visual proxy for interactions and visual encodings.

\textbf{Attribute.}
First, we encode the traffic volume passing through an intersection with the node's size.
The larger the size, the larger the traffic volume.
Second, we encode the traffic volume passing through a road with the road's width.
The wider the road, the larger the traffic volume.
Third, we encode the road's actual travel time with blue-orange color encoding, a colorblind-friendly color scheme.
The travel time for each road is divided by its respective FFTT. The minimum value of the resulting ratio is represented by blue, the maximum value by orange, and the remaining values are assigned colors through linear interpolation, with the middle value represented by light gray.
Users can hover over a road, and a tooltip pops up and shows the road's actual travel time and traffic volume.

The aforementioned design leverages existing traffic visualization techniques, aiming primarily to minimize the user's learning curve and cognitive load.

\subsubsection{Inspecting the Travel Demands}
The user can double-click a road to trigger the mode for visualizing the travel demands.
\autoref{fig:odvis}A is the snapshot after the user double-clicks the orange road.
We first extract the OD trips that pass through the clicked road.
In each node of the road network, two groups of OD trips terminate and originate there, respectively.
We employ the pie chart to visualize the quantities of these two groups.
Specifically, the pie's size encodes the total number of OD trips in the two groups.
The pink sector is for the OD trip group terminating there, and the green sector is for the OD trip group originating there.

We do not explicitly visualize the OD pairs and trips because direct depiction with lines will cause visual clutter, and the use of novel designs~\cite{DBLP:journals/cgf/ZengSJT19,deng2022multilevel,DBLP:journals/tvcg/YangDGM17} will increase the system learning cost.
Furthermore, the users in our scenario do not need to analyze the individual OD pairs and OD trips.

\subsubsection{Modifying the Road Network~\label{sec:modifications}}

Traffic planning encompasses numerous real-world countermeasures, making it impractical to enumerate them all. This study specifically focuses on roads, i.e., the edges of the road network.
We worked with experts to conclude four basic types of road-level road network modifications (\textbf{M1 to M4}).
Each modification generates a new road network, and the traffic assignment method computes the new road network state accordingly.
\begin{figure}[tb!]
  \centering 
  \includegraphics[width=\columnwidth]{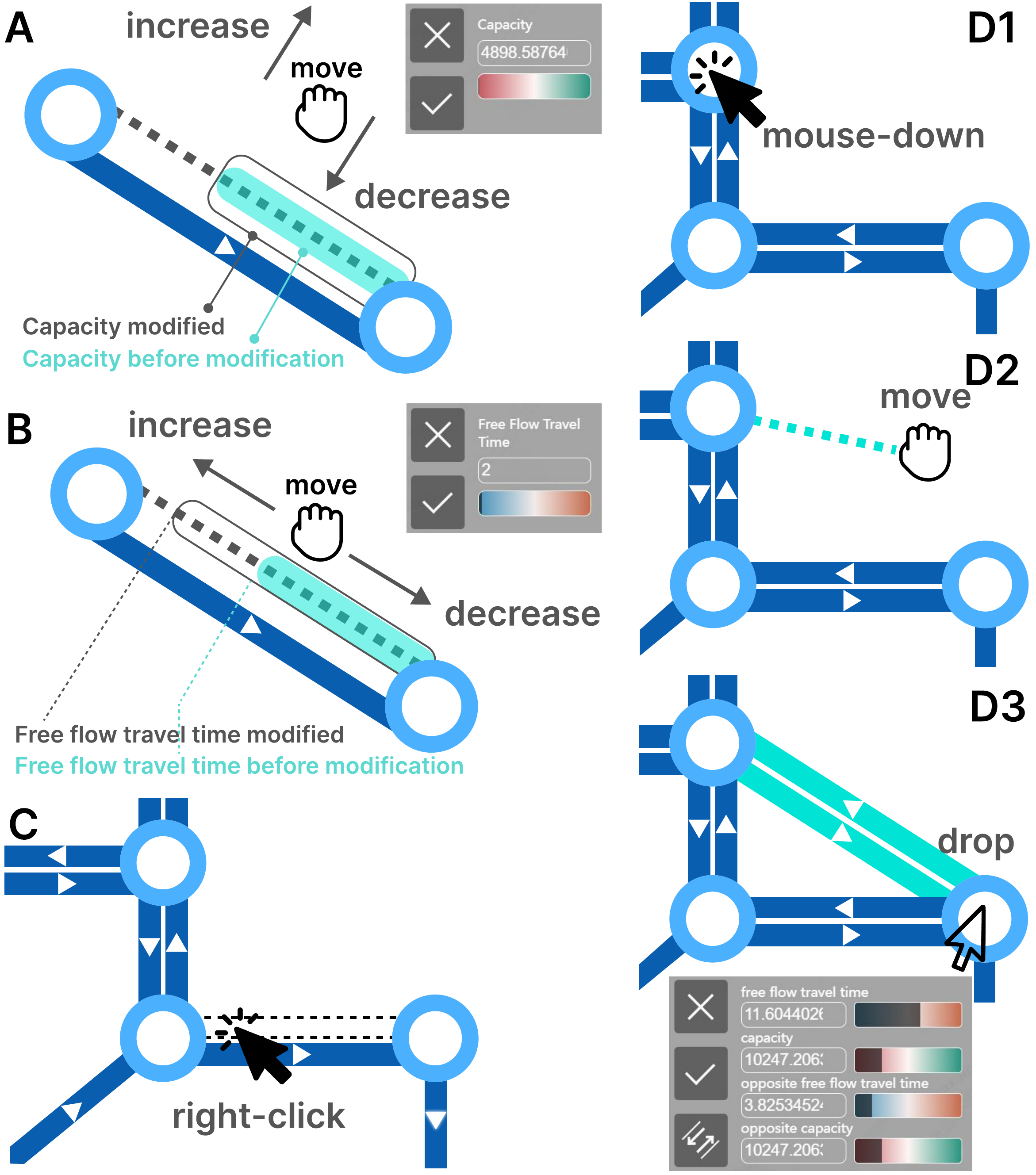}
  \caption{%
  The interactions for modifying the road network. (A) The user can expand or narrow a road to modify the road's capacity by dragging a virtual mask or inputting the value of capacity. (B) The user can improve or restrict a road to modify the free flow travel time by stretching or shortening a virtual mask or inputting the value. (C) The user can close a road by right-clicking it. (D1, D2, D3) New roads can be created between two intersections by connecting the two nodes.
  }
  \label{fig:MF}
\end{figure}

\textbf{M1: Expand/Narrow the road.}
A road can be narrowed for maintenance or if the lane is planned to be converted into a non-motorized lane.
The road can also be expanded.
These countermeasures can be considered to increase or decrease the road's capacity.
To do that, the user can first click the road and then click the ``Expand/Narrow'' icon from the popped-up panel.
Afterward, a semi-transparent mask is shown on the road, and its initial width encodes the current capacity.
The user can increase (or decrease) the capacity by moving the mouse in (or out) along the vertical direction of the road, as illustrated in \autoref{fig:MF}A.
The mask's width is updated according to the movement of the mouse.
The wider the mask, the larger the capacity.
Concurrently, a UI panel with inputs pops up for the user to specify the road's capacity precisely.

\textbf{M2: Improve/Restrict the road.}
The transportation department may improve road surface quality, reducing the free flow travel time (FFTT).
Alternatively, the department may also issue traffic control measures on the road to restrict travel speed, increasing the road's FFTT.
These countermeasures can be abstracted as the increase or decrease of the road's free flow travel time.
To do that, the user can click the ``Improve/Restrict'' icon from the popped-up panel.
Afterward, a semi-transparent mask is also shown on the road, and its initial length encodes the current FFTT.
The user can increase (or decrease) the FFTT by moving the mouse in (or out) along the road, as illustrated in \autoref{fig:MF}B.
The mask's length is updated based on the modified FFTT.
The longer the mask, the larger the FFTT.
Similar to M1, there is a UI panel that allows the user to specify the road's FFTT precisely.

\textbf{M3: Close a road.}
The transportation department may need to temporarily close a road for various reasons, such as infrastructure maintenance and traffic routing.
This countermeasure can be mapped to remove an edge in the road network.
To do that, the user only needs to right-click the road he/she wants to remove.
Such an interaction is illustrated in \autoref{fig:MF}C, where the removed road is represented by dashed lines.

\textbf{M4: Build a new road.}
Building new roads (even tunnels or overpasses) is commonly seen in urbanization to meet the rapid growth of traffic demand.
Such a countermeasure can be translated to add a new edge between two nodes.
To add an edge, the user should first mouse down on one node and drop the pointer on another node, as shown in \autoref{fig:MF}D. 
A dashed line will appear as a hint between the pointer and the starting node during this process.
By default, a two-way road is added.
The user can switch to add a one-way road by selecting the ``one-way'' icon in the panel below the \autoref{fig:MF}D3.
To determine the initial free flow travel time (FFTT), we first divide the average FFTT of existing roads by their average length.
This quotient is the average FFTT a unit length of road should have.
The initial FFTT is this quotient multiplied by the length of the newly added road.
The initial capacity is the average capacity of all other roads.
The user can modify and specify the road's FFTT and capacity via the interactions and popped-up panel described in \textbf{M1} and \textbf{M2}.

These basic modifications can be combined to form more complex modifications. For example, narrowing two-way roads simultaneously simulates a scenario where a road in one direction is under construction and requires the use of the opposite road; closing a road in one direction while expanding the opposite road simulates converting a two-way road into a one-way road.

\subsection{Comparison View}
The comparison view (\autoref{fig:teaser}B) comprises a history tree and a road-state matrix.
The history tree is designed to show the iterative modification process (\textbf{R4}), while the road-state matrix is aligned with the tree to facilitate the comparison of the effects brought about by each modification (\textbf{R5}).
Moreover, we equip this view with an indicator panel for road ranking and filtering (\textbf{R6}).

\subsubsection{Visualizing the State History}
The iterative modification process inherently exhibits a tree structure.
Each node represents a state of the road network; we call it an s-node to distinguish it from the node in a road network.
Each edge from the parent s-node to the child s-node indicates that the state of the parent s-node is modified based on the state of the parent s-node; we call it an m-edge to distinguish it from the edge in a road network.
We design the history tree (\autoref{fig:teaser}B1) to visualize the modification process.

\textbf{Structure.}
The history tree is laid horizontally.
Each s-node occupies a column with equal horizontal space.
In this way, we have sufficient screen space to visualize roads' detailed information for the corresponding state, and the matrix can be aligned with the tree by the state.
The tree is initialized as a single s-node to represent the original state of the road network before any modifications are made.
It grows to the right as the user modifies the road network.

\textbf{Optimization Metric.}
The history tree should also visualize how the road network improves or worsens during modification.
We calculate the optimization metric, $\sum_{e \in E} f^e_a t^e_a$, which is the travel time of all drivers over the network.
This metric is different from the goal introduced in equation (2), because we discard the user's satisfaction and integral term for individuals since the department pays more attention to the system efficiency rather than the individual travel experience.

Each s-node's border color encodes the optimization metric difference between the current and initial states.
The s-node of the initial state is white.
Among all the modified states, improvements regarding the metric are represented by bright blue and deteriorations by orange. The state with the maximum metric change is depicted in the darkest color, while the remaining states are assigned darkness through linear interpolation from zero to the maximum change value.
Each m-edge's color encodes the metric change between the current state and its parent (previous) state.
Blue indicates the current state improves, and orange indicates it deteriorates; the greater the change, the darker the color, following the same color scheme and scale as s-nodes.

\textbf{Cost.}
While improvements to the traffic system enhance efficiency, they also incur costs. To raise users' awareness, a sequence of icons representing these modifications is placed in the bottom-right corner of each column header, with the cumulative cost displayed above the icons.
Based on experts' suggestions, the costs for the surface road and tunnel are estimated.
Specifically, the cost is estimated at $c_r$ per kilometer for the expansion and construction of surface roads and $ c_t$ per kilometer for the construction of new tunnels. Both $c_r$ and $c_t$ have default values, 4 million and 14 million, that users can adjust.
The costs of narrowing, improving, restricting, and closing roads are negligible compared to the aforementioned modifications.
Note that actual costs depend on complex factors such as soil foundation and the surrounding environment; therefore, these estimates are rough and intended for auxiliary purposes only.

\textbf{Interaction.}
Users can select any state of the road network by clicking the s-node.
The clicked s-node will be dashed for a visual hint, and the map view will visualize the state of the corresponding road network.
When a user modifies this road network, the tree will expand by adding a new s-node to connect the s-node of this road network as its child.
In addition, users can remove any state by right-clicking the node.
Once a state is removed, its descendants will also be removed.

\subsubsection{Comparing Multiple States}
In addition to the overall optimization metric, experts need to unfold each state to perform detailed comparisons on roads.
We design a road-state matrix (\autoref{fig:teaser}B2) aligned with the history tree to support side-by-side comparison.
The matrix, together with the tree, is pannable and zoomable, holding the potential for rendering many states and roads.
For illustration, we use an example in \autoref{fig:paradox}C to demonstrate the matrix's layout and encoding.

\textbf{Layout.}
We leverage the horizontally laid-out history tree to place the matrix: each column in the matrix denotes a state, and then, each row denotes a road.
In this way, the comparison between multiple states is maintained in the context of the iterative modifications.
To enhance the spatial context, we add a mini-map in front of each row.
The road corresponding to the row is placed in the center of the mini-map and highlighted.
The mini-map design is inspired by Compass~\cite{DBLP:journals/tvcg/DengWXBZXCW22}.

\begin{figure}[tb!]
  \centering 
  \includegraphics[width=1\columnwidth]{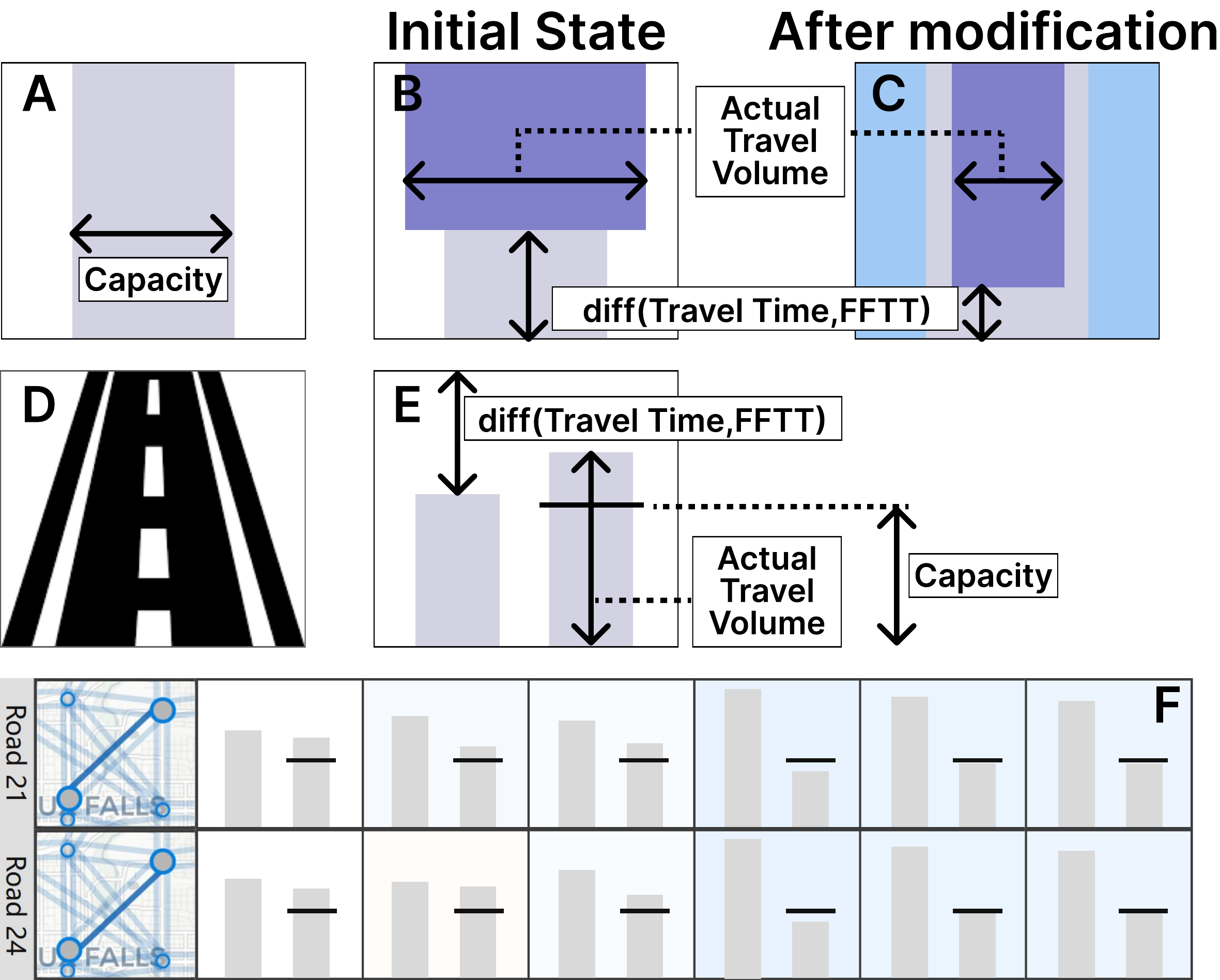}
  \caption{%
  (A, B, C) Visual design illustration of the road glyph based on (D) the road metaphor. The narrower the rectangle, the less travel flow. The taller the rectangle, the closer the travel time is to FFTT. The background color indicates whether it is improved compared to the initial state. (E, F) Alternative design.
  }
  \label{fig:roadglyph}
\end{figure}

\textbf{Encoding.}
Each road is different across states.
The road's traffic status is a four-dimensional data, characterized by the capacity $f^e$, actual traffic volume $f^e_a$, free flow travel time $t^e_f$, and actual travel time $t^e_a$.
The status should also be compared to that of the original state. 
For intuitive road comparison across states, we design a square road glyph to encode the road's status in a specific state, i.e. in each cell in the matrix.

Following the road metaphor in \autoref{fig:roadglyph}D, we first use a half-width white rectangle to indicate the capacity (\autoref{fig:roadglyph}A).
A purple rectangle is then superimposed to encode the road's status, which is illustrated in \autoref{fig:roadglyph}B or C.
The width of the purple rectangle encodes the actual traffic volume in proportion to capacity, with narrower rectangles indicating less travel flow.
The vertical distance from the purple rectangle to the bottom represents the difference between the actual travel time and the FFTT.
Taller rectangles signify closeness with FFTT and higher traffic speeds.
Finally, the glyph's background is colored orange or blue to indicate whether the travel time has increased (deterioration) or decreased (improvement) compared to the original road status.
The darkest color represents the maximum change observed across all cells. The remaining cells are shaded through linear interpolation from zero to the maximum change value, providing a gradient that encodes the change extent.

For example, assume \autoref{fig:roadglyph}B is the glyph of a road in the original state, and \autoref{fig:roadglyph}C is the glyph of this road after modification.
Initially, the actual traffic volume exceeds the capacity, as shown by the purple rectangle's width.
After modification, reduced traffic flow leads to travel times much closer to FFTT, evident from the increased height of the purple rectangle.
The blue background color indicates a reduction in travel time, highlighting the improvement.

\textbf{Encoding Justification.}
Popular visualizations for multidimensional data include dimensionality reduction, parallel coordinate plots, scatterplot matrices, and customized designs like glyphs.
Customized designs are the most suitable given the limited space.
We initially try to use the common bar-based design (\autoref{fig:roadglyph}E), in which two bars encode the attributes of travel time and volume, respectively.
Both use the length channel in the same direction, making the comparison of travel flow or travel time confusing.
Particularly, when there are many columns (\autoref{fig:roadglyph}F), only those spaced one apart are comparable.
To mitigate this issue, we adopted two orthogonal length channels: horizontal width for travel flow and vertical height for travel time, inspired by the road metaphor.
Encoding traffic volume with horizontal width provides a visual cue that as more vehicles flow onto the road, it appears increasingly likely to ``burst.''
Besides, the area of the purple rectangle is the travel time saved by each driver multiplied by the number of drivers (i.e., the traffic volume), and thus, the area can be considered the gain of the road in this road network state.

\textbf{Filtering and Ranking.}
For a large network, the roads of interest to the user (e.g., those that differ obviously between different plan candidates) will be submerged in a large number of roads in the network.
To this end, we implement an indicator panel (\autoref{fig:teaser}B3) equipped with ranking and filtering interactions to allow users to locate roads of interest quickly.
First, users select the states they are interested in by checking the box at the top of the columns.
By default, all states in the history tree are selected.
Second, 4 indicators are computed for each road (row): the averages of 1) traffic flow (i.e., actual traffic volume), 2) the ratio between flow and capacity, 3) travel time, and 4) the ratio between FFTT and travel time.
The scopes (i.e., range) of these four aspects over the selected states are also computed to characterize the road status change, resulting in 8 indicators in total.
For each indicator, a histogram will show the distribution of this indicator for all roads.
Users can filter the roads by interactively specifying a range on the histogram.
They can also rank the roads by one of the indicators via the ``ranking'' button above the histogram.

\subsubsection{Justification}
Visual comparison techniques can be divided into three strategies~\cite{DBLP:journals/ivs/GleicherAWJHR11}.

With the \textbf{superposition} strategy, road networks are plotted on the same map with a slight shift between each other.
This way is not scalable to multiple road networks since there isn't much space available on the map.
Moreover, each road's travel time and volume are hard to intuitively be encoded on the map that is already information-dense.

With the \textbf{juxtaposition} strategy, one can place road networks side-by-side.
If using \textit{small multiples of maps} where the spatial context is preserved, cross-view coordination and highlights can help compare each road across different road networks.
Again, how to encode the road's travel time and volume in each map remains challenging.

In our study, we design a road-state \textit{matrix}.
Despite the loss of spatial context, it supports a side-by-side comparison of roads and road networks at the same time.
Moreover, the cell space in the matrix layout can be leveraged to visualize the travel time and volume for detailed comparative analysis.

In the \textbf{explicit encoding}, differences are first derived and then visually exposed to users.
We also apply the explicit encoding for better comparison.
For each road network state, we compute the difference between its and the original state and encode the difference in the history tree.
For each road status, the difference between its and the original status is also visualized in the cell of the matrix.

\section{System Architecture}
TraSculptor is implemented as a web application, which is naturally separated into a server module and a client module.
The server module serves the traffic assignment model (Section \ref{Sec:model}) that computes the actual traffic volume and actual travel time of all roads and outputs them to the client part as the road network state, given a road network and OD pairs.
It is written in Python via the Flask framework.
The client module is mainly a visualization interface (Section \ref{Sec:design}), where users can access the road network state, modify the road network to improve it, and evaluate the modifications.
It is written in TypeScript via the Vue3 framework.
The server and client modules are tightly integrated, supporting an iterative workflow.
For a road network, whether original or modified, the client module will receive the computed road network state from the server module.
The state will be visualized in the interface for users to access, modify, and improve it further.

\section{Evaluation}
This section introduces a usage scenario, a case study, and expert interviews that were performed to evaluate TraSculptor.

\begin{figure}[tb!]
  \centering 
  \includegraphics[width=\columnwidth]{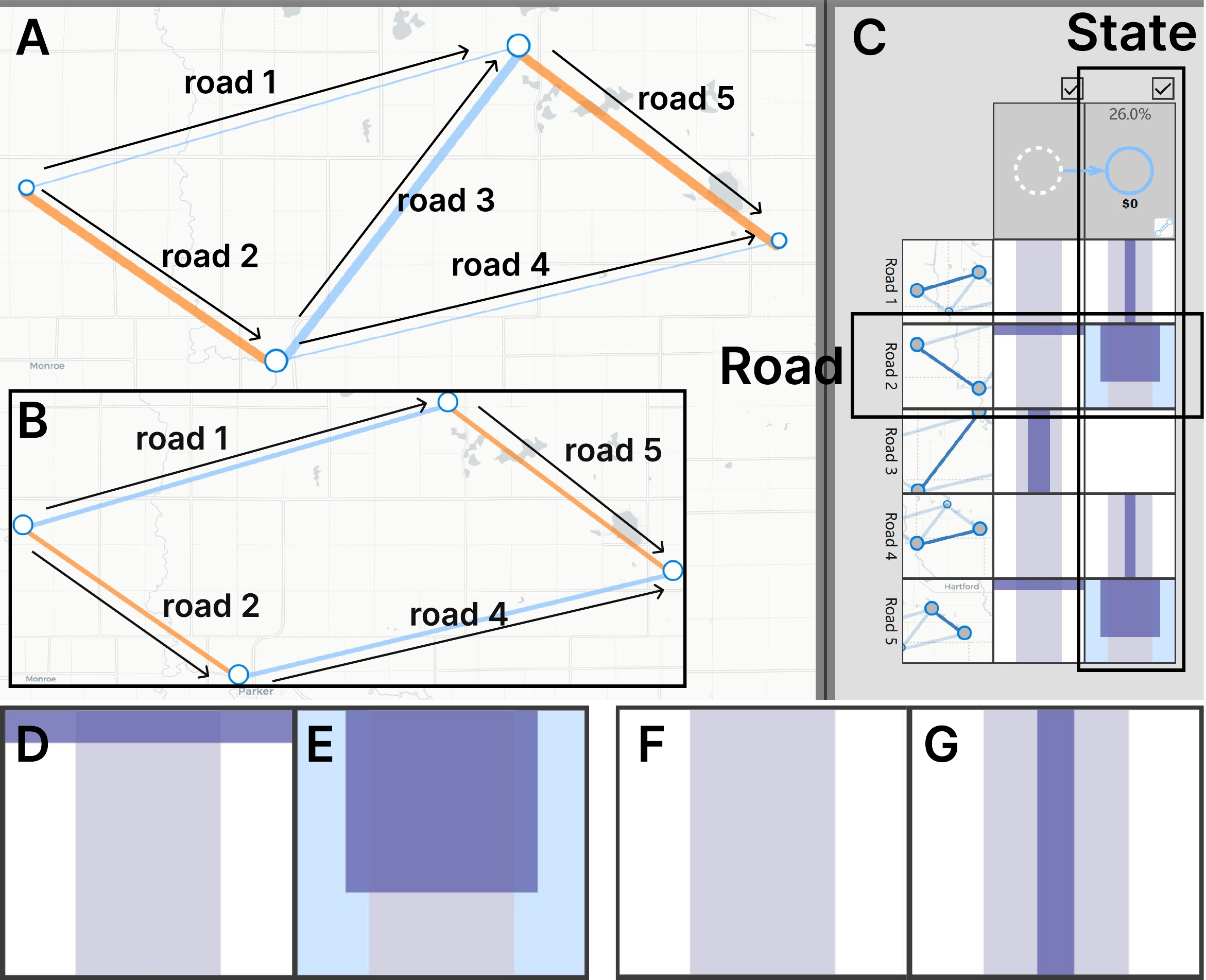}
  \caption{%
  The usage scenario of Braess's paradox. (A) The illustrative road network. (B) The road network after road 3 is closed. (C) The comparison between the two states before and after modification, respectively. (D, E) The glyphs of roads 2 and 5 in the comparison view. (F, G) The glyphs of roads 1 and 4 in the comparison view.
  }
  \label{fig:paradox}
\end{figure}

\subsection{Usage Scenario}
It is not wise to build roads blindly.
Sometimes, building a new road may increase travelers' travel time, while closing a road may reduce the travel time.
Such cases are considered the Braess paradox~\cite{braess1968paradoxon,steinberg1983prevalence} and have been found in Seoul, New York City, and Stuttgart\footnote{\url{https://en.wikipedia.org/wiki/Braess\%27s\_paradox}}. 

To test whether our method can detect such cases, we used an illustrative road network that is widely used to showcase the Braess paradox. 
The illustrative road network (\autoref{fig:paradox}A) comprises 4 intersections and 5 roads (i.e., roads 1-5).
There is one OD pair with a travel demand of 1000 from the leftmost node (i.e., the origin) to the rightmost node (i.e., the destination).
Three paths exist for this OD pair, i.e., path 1-5 (the abbreviation of road 1 and then road 5), path 2-3-5, and path 2-4.
Please refer to the appendix for the capacity and FFTT of every road.

\textbf{Intelligence.}
The initial state of the road network is visualized as \autoref{fig:paradox}A. 
Roads 2, 3, and 5 are much wider than roads 1 and 4, because most drivers chose the path 2-3-5.
In this path, roads 2 and 5 are orange.
The long travel times due to the large amount of traffic (far exceeding the road's capacity) that pours into these two roads.
Nonetheless, the road 3 is blue with a short travel time due to its large capacity.

The comparison view provides detailed evidence.
\autoref{fig:paradox}D is an enlarged version of the glyph of road 2 or road 5 in \autoref{fig:paradox}C.
On the one hand, the purple rectangle is very wide, which means that actual traffic far exceeds capacity.
On the other hand, the purple rectangle is very short, which means that the actual travel time is much greater than the FFTT.
\autoref{fig:paradox}F enlarges the glyph of road 1 or road 4 in \autoref{fig:paradox}C.
It shows that few vehicles are on the road 1 or the road 4.



In sum, the overall travel times of the drivers are very high.
If some drivers can drive to the road 1, the congestion on the road 2 will be greatly alleviated.
However, road 3, which has a large capacity and a small FFTT, attracts a large number of drivers to road 2.
Then, the next task is to verify whether closing the road 3 will improve the efficiency of the transportation system.
With the tooltips, we can check the travel time for the other two paths, which is 191.9, while the travel time for the current path is around 168.8.
Thus, no drivers will choose the other two paths because most drivers are ``selfish'' to choose the ``shortest'' path.
Only a few drivers choose the path 1-5 or 2-4 since we applied the USE traffic assignment method that assumes drivers don't always have an accurate sense of road status.

\textbf{Design.}
We right-click the road 3 to remove it, obtaining the latest state of the illustrative road network (\autoref{fig:paradox}B).
Compared with \autoref{fig:paradox}A, the roads 2 and 5 get narrower, while the roads 1 and 4 become wider.
These roads have nearly the same width, indicating that traffic demands are evenly distributed between the remaining two paths.
Vehicles on both paths have the same travel time, and drivers do not prefer one path over the other.

From the comparison view (\autoref{fig:paradox}C), we can see the status changes of the four roads.
On the one hand, the traffic volume on roads 2 and 5 decreases, and the travel time decreases significantly (\autoref{fig:paradox}D and E).
On the other hand, the traffic volume on roads 1 and 4 increases but is still below capacity (\autoref{fig:paradox}F and G).
Vehicles on the roads can quickly pass through the roads.
Taken together, the network has been improved greatly with 26\% improvement.
The travel time of each vehicle can be checked via the tooltip as 124.5.

\textbf{Summary.}
In this usage scenario, we illustrate how TraSculptor assists users in accessing the road network, simulating the countermeasures with intentions directly on the road network, and evaluating the countermeasure with comparative analyses.
This scenario builds upon the interesting Braess's paradox that is widely used by researchers in the transportation domain.
Particularly, drivers' choices are ``selfish,'' and thus, the existence of road 3 will cause one driver after another to take the path (i.e., path 2-3-5) with the shortest travel time from their perspective.
Even when all drivers take the path 2-3-5, the other two paths are still at a travel time disadvantage, so no one switches until road 3 is closed.
Please refer to the Wikipedia page (in the footnote) for more information about the Braess's paradox.

\begin{figure}[tb!]
  \centering 
  \includegraphics[width=\columnwidth]{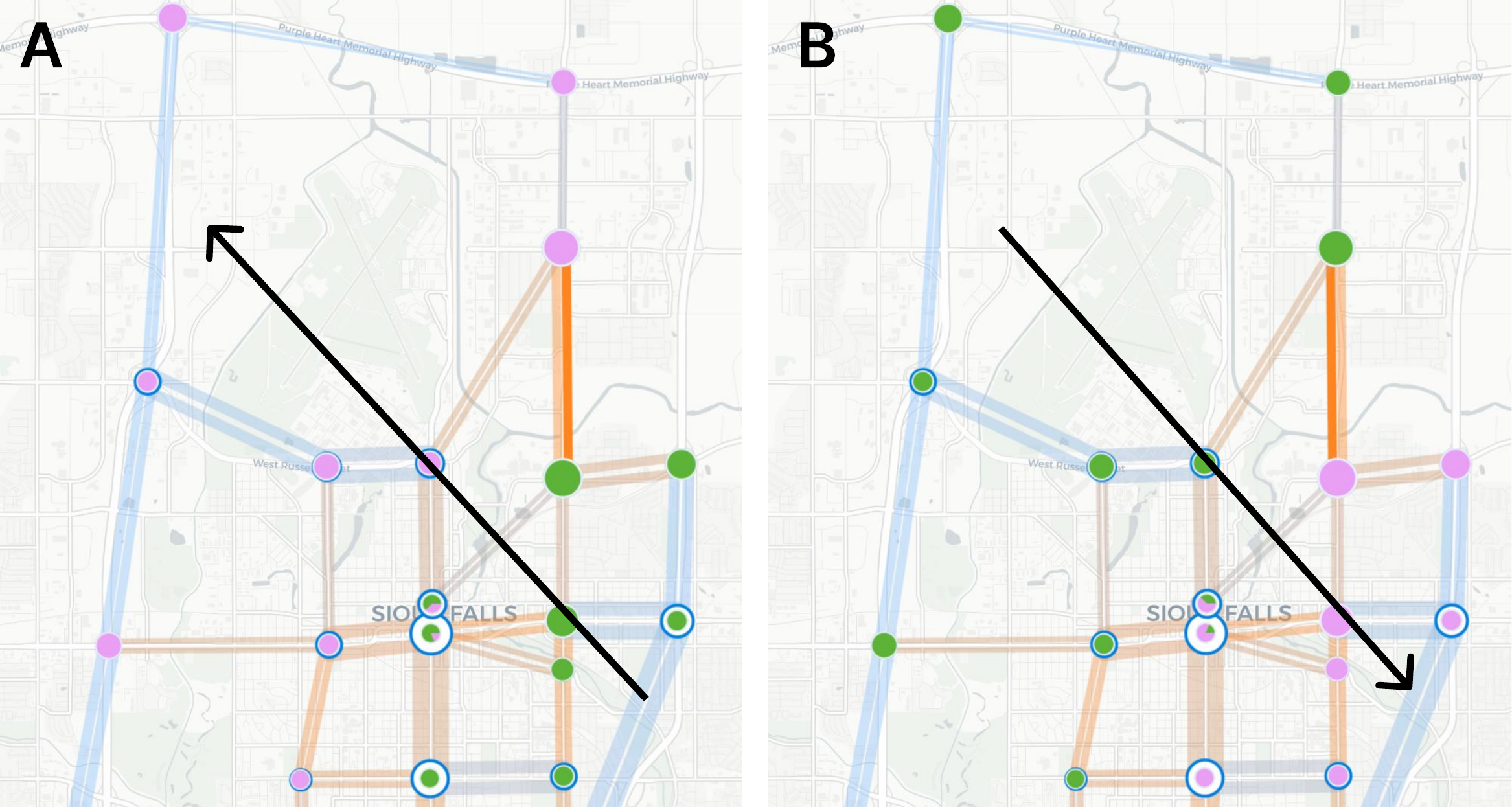}
  \caption{%
  (A, B) The distribution of the OD trips on the most congested roads, respectively. The pink and green sectors in the pie chart encode traffic volumes terminating and originating there, respectively. The black arrow indicates the general movement direction of the OD trips.
  }
  \label{fig:odvis}
\end{figure}

\subsection{Case Study}
\textbf{Procedure.} We invited the three domain experts (EA, EB, and EC) and hosted a case study section.
Since the two experts were in different cities, this session was conducted through a hybrid online and offline meeting.
The system is deployed via a cloud server, and every expert can use the system personally during the meeting.
For synchronization, one expert shared the screen and operated the system for the other two.
This section includes a warming-up tutorial where we introduced the visual encodings and interactions following the usage scenario above and ensured that all experts fully understood the system.
The case study section lasted for one hour.

\textbf{Dataset.}
Due to data sensitivity and privacy concerns, private datasets from real business scenarios were unavailable for the case study.
Using a public dataset ensures the reproducibility of our results.
In particular, we selected the Sioux Falls dataset~\cite{leblanc1975efficient,leblanc1975algorithm} due to its established recognition and widespread adoption within the transportation research community~\cite{wang2013global,di2020reversible,yin2022simulation}.
The network comprises 24 intersections and 76 roads, with 552 OD pairs and 360,600 OD trips.
Although synthetic, it comprises a real road network in Sioux Falls, South Dakota, USA, capturing typical urban characteristics and serving as a good benchmark.
Its extensive use in academic research qualifies it as a ``real'' dataset.
While considering alternative public datasets\footnote{\url{https://github.com/bstabler/TransportationNetworks}}, we found them unsuitable due to a lack of geographic coordinates or excessive runtime requirements.
Since our contribution focuses on a visual decision-making framework, the scalability issue caused by computation was not addressed in this study.
Please refer to Section \ref{sec:discuss} for the discussion on the scalability.

\textbf{Intelligence.}
After feeding this dataset into the traffic assignment method, the expert obtained the initial state shown in \autoref{fig:teaser}A1.
The inner roads are mostly orange; in contrast, the outer roads tend to appear blue.
The expert said that the center of the area is busier, and many vehicles need to go there or pass through it.
The expert was very interested in the most congested roads (i.e., roads 16 and 19) with the reddest color (\autoref{fig:teaser}A2).
To determine the reason for the bad road status, the expert double-clicked these two roads one by one.
\autoref{fig:odvis}A and B showed the origins and destinations of the OD trips passing through these two roads, respectively.
It was obvious that many OD trips between the southeast and northwest areas pass through these two roads.
Also, because of this, the roads of \autoref{fig:teaser}A3 were orange and very congested.

\begin{figure}[tb!]
  \centering 
  \includegraphics[width=\columnwidth]{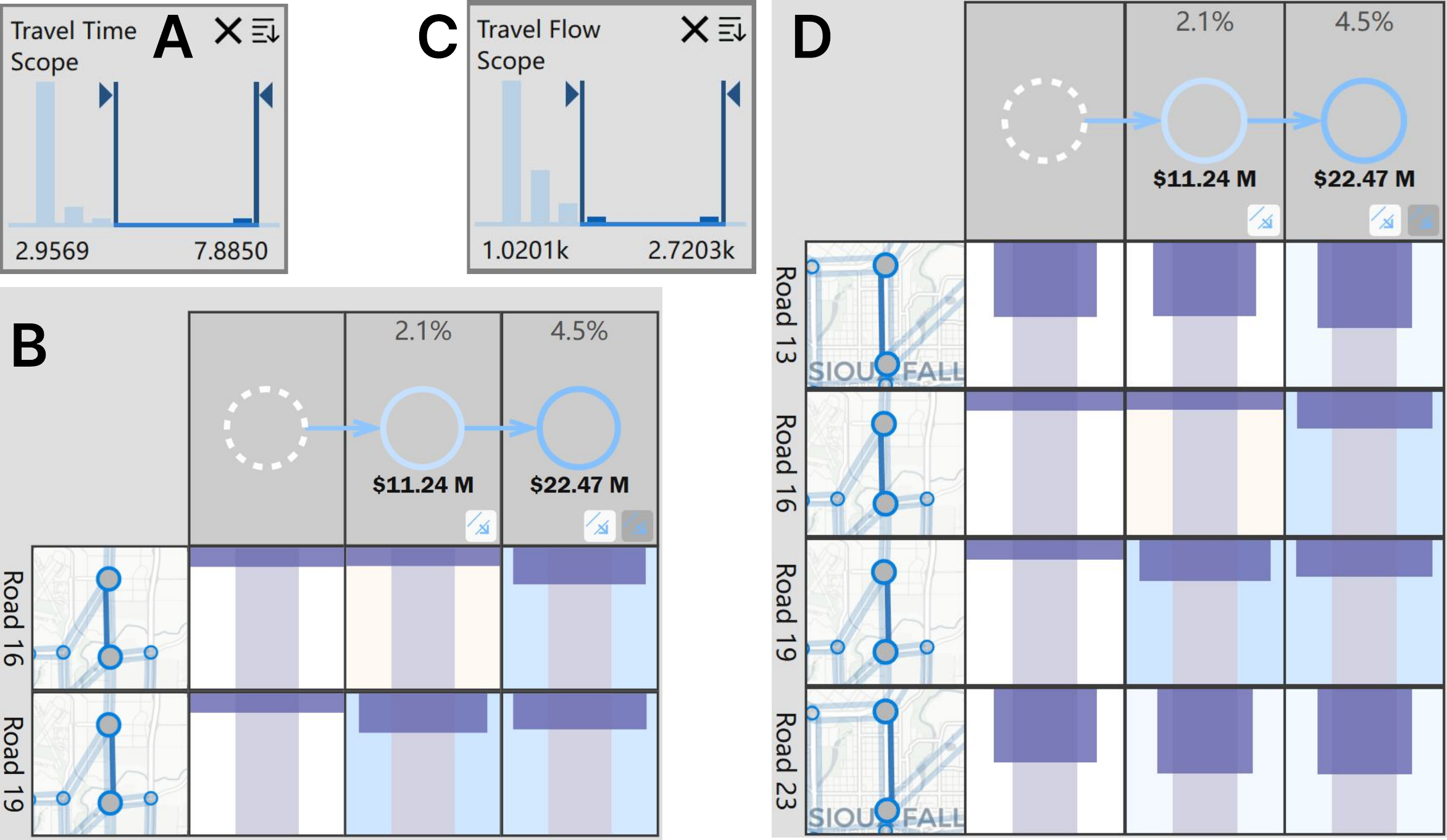}
  \caption{%
  Round I of improving Sioux Falls road network by expanding roads 16 and 19. After filtering the roads with large changes in the (A) travel time, (B) only two roads were left. After filtering the roads with large changes in the (C) traffic flow (volume), (D) four roads were left.
  }
  \label{fig:expandroads}
\end{figure}

\textbf{Design and Choice: Round I.}
The most straightforward countermeasure is expanding or improving roads 16 and 19.
Using the provided interactions, the expert first expanded them to 1.5 times their original width.
The capacities of both roads were changed from 4800 to about 7200.
However, such a countermeasure only led to a 4.5\% improvement regarding the optimization metric, which was very limited.
Moreover, the improvement mostly comes from the two modified roads.
Specifically, the expert filtered the road with large changes in the travel time (\autoref{fig:expandroads}A) and obtained only two roads (\autoref{fig:expandroads}B).
He also filtered the road with large changes in the traffic volume (\autoref{fig:expandroads}C), and only four roads were left (\autoref{fig:expandroads}D).
In sum, the expert considered that expanding/improving individual roads scratches the surface and thus sought other options.

\textbf{Design and Choice: Round II.}
Inspired by the visualized OD demand (\autoref{fig:odvis}), the expert tried to build new roads (or tunnels) to serve the traffic demands.
The expert performed three modifications based on the initial road network, as shown in P1, P2, and P3 of \autoref{fig:teaser}A, respectively.
P1 corresponded to the fourth node in the history tree (\autoref{fig:teaser}B1), where a two-way road (two roads form a two-way road) was added.
P2 and P3 corresponded to the fifth and sixth nodes in the history tree, respectively, where two-way tunnels were added with higher construction costs.
In the history tree (\autoref{fig:teaser}B1), these three columns had greener background colors than the third column, which corresponded to larger improvements (11.6\%, 16.9\%, and 16.9\%).
The expert was very satisfied with the road-building countermeasures and will choose the most ideal one among them.

\begin{figure*}[tb!]
  \centering 
  \includegraphics[width=2\columnwidth]{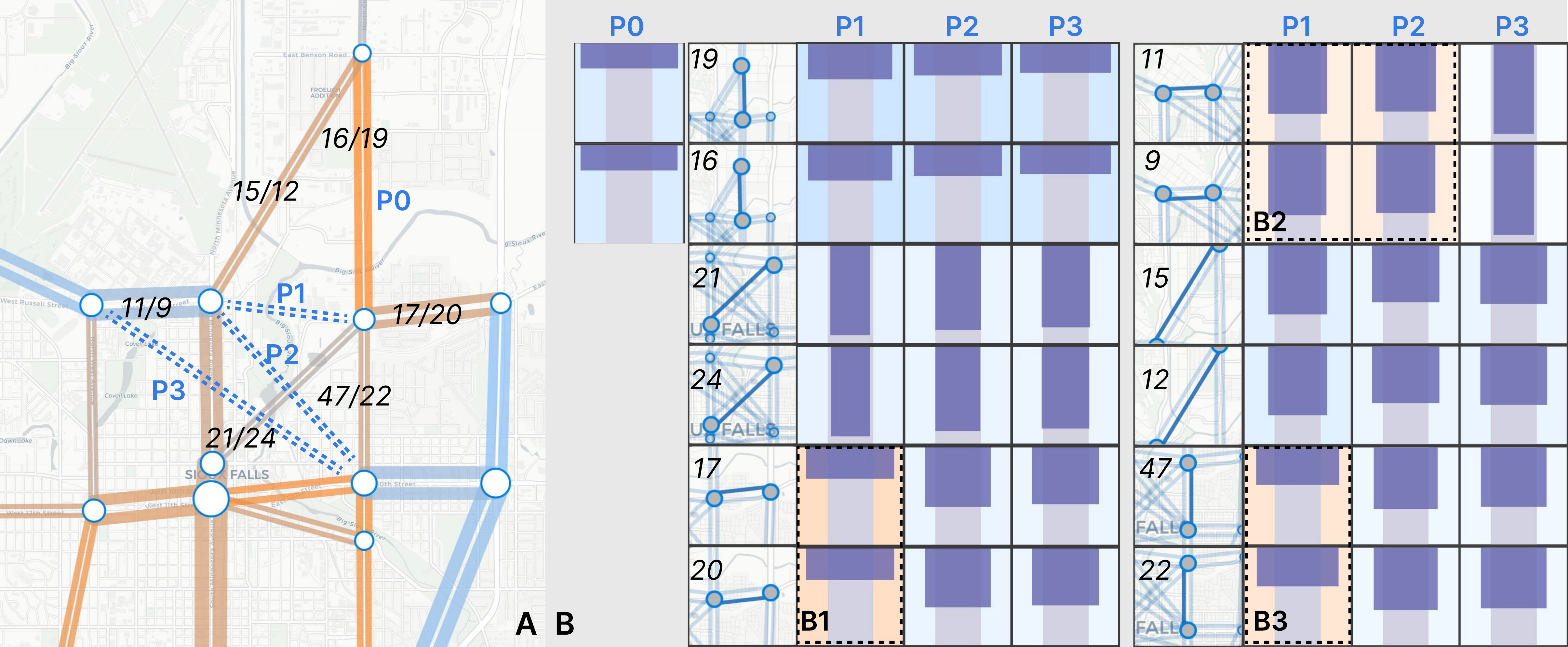}
  \caption{%
  Round II of improving Sioux Falls road network by building new roads. (A) The map view mainly showed the spatial context of (P1, P2, and P3) three newly built roads. The roads that will be affected were annotated. (B) The comparison view mainly showed detailed road status changes across the modifications of P1, P2, and P3. (B1, B2, and B3) The roads became worse compared to their status in the original road network.
  }
  \label{fig:comparison}
\end{figure*}

\textit{Filtering and Ranking.}
To compare the three modifications, the expert first deselected the second and third columns.
The expert first chose to filter the roads.
From the indicator panel (\autoref{fig:teaser}B4), it can be seen that the traffic volume of many roads reaches or even exceeds the capacity.
The expert filtered these roads by adjusting the range on the histogram of the traffic flow ratio and ranked the roads by the scope of the traffic flow ratio in descending order.
The expert obtained the comparison view shown in \autoref{fig:teaser}B2.
The top 12 roads in this view were carefully analyzed, illustrated in \autoref{fig:comparison}.
\autoref{fig:comparison}A is the road network on the map while \autoref{fig:comparison}B is the results of the comparison view.
For illustration, we also denote the countermeasure of improving roads as P0.

\textit{Analyzing roads 16 and 19.}
The first two rows were the two most congested roads, i.e., the roads 16 and 19.
On these two roads, P1 led to the largest improvement, followed by P2, and then P3.
The expert explained that the new roads in P1 were closest to the two roads and could serve many travel demands that needed to pass through the two roads before.
The improvement caused by P3 is comparable to that caused by P0.
The expert said, ``\textit{Building new roads can provide comparable improvements to P0 on roads 16 and 19, and have a more positive impact on the road network rather than a single road.}''

\textit{Analyzing other roads.}
The expert also analyzed the improvements of P1, P2, and P3 on other roads.
First, P1 negatively impacts roads 17 and 20, roads 11 and 9, and roads 47 and 22, indicated by the orange background of glyphs below P1 in \autoref{fig:comparison}B.
The expert explained that these roads were directly connected to the new roads (\autoref{fig:comparison}A), and most of the traffic flow on the new roads needed to pass through these roads, which would make these roads congested.
Particularly, roads 17 and 20 and roads 47 and 22 already tend to crowd as they were light orange-brown.
Second, P2 had negative impacts on roads 11 and 9, indicated by the orange background of glyphs below P2 in \autoref{fig:comparison}B.
It was also because the newly built tunnel was connected to roads 11 and 9 and attracted a part of traffic demands.
Third, the glyphs below P3 in \autoref{fig:comparison}B were blue, indicating that P3 improved all these roads.

The expert preferred P2 because
1) the construction cost of P3 was the highest as the tunnel was long and went through two roads (P3 with 23.77 million and P2 with 15.07 million);
2) P1 made some roads that are already unobstructed more unobstructed (\autoref{fig:comparison}B1 and B3);
3) although P2 had a negative impact on roads 11 and 9, the traffic volume there had not seriously exceeded the capacities, and the road conditions were still acceptable (\autoref{fig:comparison}B2).

\subsection{Expert Interview}
After the case study, we interview the experts one by one to collect their feedback. The feedback is summarized as follows:

\textbf{Intuitiveness.}
All experts confirmed the intuitiveness of the interactions and visualizations.
First, the road network visualization draws on the visualization in common traffic simulation platforms.
Second, the interactions for network modification can be naturally related to real-world traffic countermeasures.
EA spoke highly, ``\textit{it is intuitive that the interactions can be directly performed on the road network rather than with typing inputs in another separate view}.''
Third, the comparison view organizes multiple roads in multiple road networks well for comparison.
Furthermore, ``\textit{the glyph is intuitive because the width is often used to encode traffic, and the area and color are easy to perceive},'' commented EB.

\textbf{Usability.}
All experts considered TraSculptor to be highly useful.
TraSculptor empowers the what-if analysis of the road network, which is important in the decision-making of traffic planning.
EC commented that ``\textit{we can perform a variety of what-if analysis}.''
TraSculptor visualizes necessary information of the road network and provides them numerically, such as travel times, volumes, and improvements.
EB said, ``\textit{the information makes decisions more reliable.}''

\textbf{Significance.}
All experts thought that the proposed interactions and visualizations could make up for the shortcomings of existing traffic simulation platforms in interactive and iterative traffic planning.
In addition, EA and EC commented on the significance of TraSculptor from their own perspective based on their daily work.

EA, a university professor, said TraSculptor could be used for better education.
Students majoring in transportation need to use traffic planning platforms, such as TransCAD, to perform traffic assignment experiments.
However, existing traffic planning platforms currently cannot intuitively visualize changes in road information, i.e., traffic volume and travel time.
For example, understanding Braess's paradox requires visualization to display road information in the two states of the road network.
TraSculptor fills the gap and ``\textit{can aid in the teaching of the traffic assignment theory and methods}.''
EC, who is conducting research on the automatic optimization of road networks, hoped that we could provide flexible APIs for him and even other scholars in the field of transportation planning.
He believed TraSculptor could promote the research in his field:
``\textit{with TraSculptor, we can intuitively and easily analyze the optimization results and compare them.}''
\section{Discussion~\label{sec:discuss}}

\textbf{Lessons Learned.}
We would like to share three lessons with fellow visual analytics and visualization researchers.

\textit{Problem Seeking.} Many visual analytics studies focus on analytical tasks that are not yet supported, such as in fields of sports science~\cite{DBLP:journals/cgf/PerinVSSWC18}, urban science~\cite{TrafficSurvey,DBLP:journals/cvm/DengWLTXW23}, and social media~\cite{DBLP:journals/tmm/WuCGTK16}.
Through this collaboration, we learned that even if an analysis task is already supported by mature and commercial software platforms, many aspects can still be improved through visual analytics and human-computer interaction.

\textit{Problem Characterization.}
A fundamental aspect of a design study is to characterize the domain problem.
To summarize the requirements, we follow the expert interview guideline recently proposed by Cibulski et al.~\cite{cibulski2022supporting}.
Based on our experience, adhering to this guideline allows us to gather overall and substantial requirements within approximately two hours of interviews.
Specifically, involving experts to review their analysis process (i.e., ``Unstructured Incident Recall'' in the guideline~\cite{cibulski2022supporting}), from loading data to gaining results, enabled us to delineate their workflow into intelligence, design, and decision stages.
The intelligence stage encompasses traffic status and travel demand analysis; the design stage involves interactive editing of the road network; the decision stage focuses on comparing different simulation results.
Following this established framework, subsequent meetings iteratively refine the details, including the types of interactions required and the dimensions of simulation results that necessitate comparison.

\textit{Network Comparison.}
Decomposing each network into edges and organizing multiple networks into a table effectively supports side-by-side comparisons, as evident by the comparison view in TraSculptor.
This way is particularly beneficial when each edge involves multiple dimensions of attributes, as the table-based organization allows for these attributes to be fully encoded within each cell.
In this way, our design facilitates multiple network comparisons that align with the history tree for analytical provenance, addressing the challenges associated with comparing the attributes of multiple networks and the difficulty of integrating network comparison with provenance visualization.
A potential risk is that overall network structure comparisons may become less apparent, although empty cells in the table indicate missing edges.
More research efforts are required to support both structural and attribute-based comparisons effectively.

\textbf{Scalability.}
The system's scalability can be discussed from the model and visualization perspectives, respectively.
Addressing the scalability of traffic assignment models has been a long-standing challenge.
For example, traffic assignment on the Austin network, a large-scale road network, demands approximately 30 minutes of runtime~\cite{JAFARI2017270}. Although techniques such as network aggregation~\cite{raadsen2020aggregation} and partitioning~\cite{yahia2018network, JAFARI2017270} can expedite these models, the extent of acceleration is often limited, and the availability of open-source implementations remains sparse.
Integrating such accelerated models requires substantial development effort but leads to only limited runtime performance gains.
We are unable to test the visualization scalability on large-scale road networks.
One potential research direction is investigating progressive approximation and steerable early stops (like BNVA~\cite{DBLP:journals/tvcg/WengZDMBZXW21}) to ensure seamless visual analytics.

Nonetheless, our visualizations have the potential to handle large-scale road networks.
First, the built-in zoom and pan interactions of the map view allow users to explore the road network and assess traffic conditions.
For enhanced scalability in exploration, spatial simplification techniques~\cite{DBLP:journals/tvcg/LandesbergerBRA16} and level-of-detail rendering can be integrated to visualize large-scale traffic data better.
Second, roads of interest—such as those that are highly congested or that exhibit significant differences across candidate strategies—can be quickly identified using the comparison view's filtering and ranking functionalities.
In real-world scenarios, incremental traffic planning (i.e., upgrading an existing road network) typically affects only localized areas, as a road is not affected by another road being repaired tens of kilometers away.
Thus, the number of roads requiring analysis is typically limited, enabling our visualizations to support focused examination effectively.
We use TraSculptor to analyze another larger road network with hundreds of roads, with the aim of supporting the above discussion.
Due to the limited space, we place this usage scenario in Appendix C.

\textbf{Limitations and Future Work.}
Informed and interactive road traffic planning still necessitates future work from the following aspects:

\textit{Dynamic Traffic Demands.}
Currently, the traffic assignment only considers static traffic demands.
In a real scenario, the traffic demands are time-varying, for example, leading to morning and evening rush hours.
However, designing a decision-making framework that incorporates temporal variation is rather challenging.
The spatial visualization of traffic states in the map view and comparative visualization in the comparison view need to be extended to be spatiotemporal versions by incorporating the time dimension.
In the future, we will deepen our cooperation with domain experts and study visual analytics solutions for traffic planning and decision-making based on dynamic demands.

\textit{Inclusion of Additional Decision-oriented Data.} Currently, the system focuses on the network and flow aspects of the transportation network.
For planning and deciding on reconstructions, experts should also consider material data, city development plan data, the age and material of existing roads, and so on.
These data may help to trade off multi-criteria decision problems and evaluate feasibility in addition to flow improvements.
For example, \textit{is a tunnel, elevated road, or surface road more suitable?}
These data could be included using glyphs or extended network visualization techniques.

\textit{Comprehensive Interactions.}
This study focuses exclusively on road-level traffic planning countermeasures and the corresponding edge-based interactions. 
Future work must account for a broader range of countermeasures.
Intersection-level countermeasures require node-based interactions. An example would be introducing a new road from an intersection to the midpoint of an existing road, which requires adding a node at the center of an existing edge.
Topological-level countermeasures involve implicit interactions, such as restricting vehicle movements from one road to another at a specific intersection through simultaneous manipulation of both nodes and edges.
Designing intuitive interactions will follow the same principle, that is, allowing experts to interact directly with the road network in a manner consistent with their intrinsic understanding.

\section{Conclusion}
This study presents a visual analytics approach, TraSculptor, to assist traffic planners in designing better road networks.
By characterizing the domain requirements, that is, experts need to go through the iterative road network design and simulation process, we designed flexible interactions for road network modifications, a history tree that supports iterative design process tracking, and a comparative visualization for comparing multiple road network states of simulation results during the design process.
By streamlining the iterative design process, TraSculptor empowers planners to make more informed choices, ultimately leading to the development of more efficient and sustainable urban transportation systems.

\bibliographystyle{IEEEtran}
\bibliography{template}

\begin{thebibliography}{10}
\providecommand{\url}[1]{#1}
\csname url@samestyle\endcsname
\providecommand{\newblock}{\relax}
\providecommand{\bibinfo}[2]{#2}
\providecommand{\BIBentrySTDinterwordspacing}{\spaceskip=0pt\relax}
\providecommand{\BIBentryALTinterwordstretchfactor}{4}
\providecommand{\BIBentryALTinterwordspacing}{\spaceskip=\fontdimen2\font plus
\BIBentryALTinterwordstretchfactor\fontdimen3\font minus \fontdimen4\font\relax}
\providecommand{\BIBforeignlanguage}[2]{{%
\expandafter\ifx\csname l@#1\endcsname\relax
\typeout{** WARNING: IEEEtran.bst: No hyphenation pattern has been}%
\typeout{** loaded for the language `#1'. Using the pattern for}%
\typeout{** the default language instead.}%
\else
\language=\csname l@#1\endcsname
\fi
#2}}
\providecommand{\BIBdecl}{\relax}
\BIBdecl

\bibitem{gao2024evaluation}
S.~Gao, Q.~Ran, Z.~Su, L.~Wang, W.~Ma, and R.~Hao, ``Evaluation system for urban traffic intelligence based on travel experiences: A sentiment analysis approach,'' \emph{Transportation Research Part A: Policy and Practice}, vol. 187, p. 104170, 2024.

\bibitem{merchan2020quantifying}
D.~Merchan, M.~Winkenbach, and A.~Snoeck, ``Quantifying the impact of urban road networks on the efficiency of local trips,'' \emph{Transportation Research Part A: Policy and Practice}, vol. 135, pp. 38--62, 2020.

\bibitem{antunes2003accessibility}
A.~Antunes, A.~Seco, and N.~Pinto, ``An accessibility--maximization approach to road network planning,'' \emph{Computer-Aided Civil and Infrastructure Engineering}, vol.~18, no.~3, pp. 224--240, 2003.

\bibitem{meng2008general}
Q.~Meng, W.~H. Lam, and L.~Yang, ``General stochastic user equilibrium traffic assignment problem with link capacity constraints,'' \emph{Journal of Advanced Transportation}, vol.~42, no.~4, pp. 429--465, 2008.

\bibitem{ccolak2016understanding}
S.~{\c{C}}olak, A.~Lima, and M.~C. Gonz{\'a}lez, ``Understanding congested travel in urban areas,'' \emph{Nature Communications}, vol.~7, no.~1, p. 10793, 2016.

\bibitem{LIU2020102939}
L.~Liu, M.~Zhang, and T.~Xu, ``A conceptual framework and implementation tool for land use planning for corridor transit oriented development,'' \emph{Cities}, vol. 107, p. 102939, 2020.

\bibitem{MATSim}
S.~A. O.~M. Tanvi~Maheshwari, Pieter~Fourie and K.~W. Axhausen, ``Iterative urban design and transport simulation using sketch matsim,'' \emph{Journal of Urban Design}, vol.~29, no.~2, pp. 184--207, 2024.

\bibitem{DBLP:series/lncs/KeimAFGKM08}
D.~A. Keim, G.~L. Andrienko, J.~Fekete, C.~G{\"{o}}rg, J.~Kohlhammer, and G.~Melan{\c{c}}on, ``Visual analytics: Definition, process, and challenges,'' in \emph{Information Visualization - Human-Centered Issues and Perspectives}, ser. Lecture Notes in Computer Science, 2008, vol. 4950, pp. 154--175.

\bibitem{DBLP:journals/vc/LiuCWL14}
S.~Liu, W.~Cui, Y.~Wu, and M.~Liu, ``A survey on information visualization: {R}ecent advances and challenges,'' \emph{The Visual Computer}, vol.~30, no.~12, pp. 1373--1393, 2014.

\bibitem{DBLP:journals/cvm/DengWLTXW23}
Z.~Deng, D.~Weng, S.~Liu, Y.~Tian, M.~Xu, and Y.~Wu, ``A survey of urban visual analytics: Advances and future directions,'' \emph{Computational Visual Media}, vol.~9, no.~1, pp. 3--39, 2023.

\bibitem{NEALE1997441}
D.~C. Neale and J.~M. Carroll, ``Chapter 20 - the role of metaphors in user interface design,'' in \emph{Handbook of Human-Computer Interaction (Second Edition)}, second edition~ed.\hskip 1em plus 0.5em minus 0.4em\relax Amsterdam: North-Holland, 1997, pp. 441--462.

\bibitem{Maa1992}
S.~Maa{\ss} and H.~Oberquelle, ``Perspectives and metaphors for human-computer interaction,'' in \emph{Software development and reality construction}.\hskip 1em plus 0.5em minus 0.4em\relax Springer, 1992, pp. 233--251.

\bibitem{DBLP:journals/tvcg/WangWSZLFSDC18}
Y.~Wang, Y.~Wang, Y.~Sun, L.~Zhu, K.~Lu, C.~Fu, M.~Sedlmair, O.~Deussen, and B.~Chen, ``Revisiting stress majorization as a unified framework for interactive constrained graph visualization,'' \emph{{IEEE Trans. Vis. Comput. Graph.}}, vol.~24, no.~1, pp. 489--499, 2018.

\bibitem{DBLP:journals/tvcg/PanCZZZZCFW21}
J.~Pan, W.~Chen, X.~Zhao, S.~Zhou, W.~Zeng, M.~Zhu, J.~Chen, S.~Fu, and Y.~Wu, ``Exemplar-based layout fine-tuning for node-link diagrams,'' \emph{{IEEE Trans. Vis. Comput. Graph.}}, vol.~27, no.~2, pp. 1655--1665, 2021.

\bibitem{DBLP:journals/informatics/EichnerGST16}
C.~Eichner, S.~Gladisch, H.~Schumann, and C.~Tominski, ``Direct visual editing of node attributes in graphs,'' \emph{Informatics}, vol.~3, no.~4, p.~17, 2016.

\bibitem{DBLP:journals/tvcg/DengWXBZXCW22}
Z.~Deng, D.~Weng, X.~Xie, J.~Bao, Y.~Zheng, M.~Xu, W.~Chen, and Y.~Wu, ``Compass: {T}owards better causal analysis of urban time series,'' \emph{{IEEE Trans. Vis. Comput. Graph.}}, vol.~28, no.~1, pp. 1051--1061, 2022.

\bibitem{deng2022multilevel}
Z.~Deng, S.~Chen, X.~Xie, G.~Sun, M.~Xu, D.~Weng, and Y.~Wu, ``Multilevel visual analysis of aggregate geo-networks,'' \emph{{IEEE Trans. Vis. Comput. Graph.}}, 2022.

\bibitem{DBLP:journals/tvcg/WangLYZW13}
Z.~Wang, M.~Lu, X.~Yuan, J.~Zhang, and H.~van~de Wetering, ``Visual traffic jam analysis based on trajectory data,'' \emph{{IEEE Trans. Vis. Comput. Graph.}}, vol.~19, no.~12, pp. 2159--2168, 2013.

\bibitem{kotusevski2009review}
G.~Kotusevski and K.~Hawick, ``A review of traffic simulation software,'' 2009.

\bibitem{moller2014introduction}
D.~P. M{\"o}ller, ``Introduction to transportation analysis, modeling and simulation,'' \emph{Simulation Foundations, Methods and Applications. London: Springer London}, 2014.

\bibitem{su2023hierarchical}
Z.~Su, A.~H. Chow, C.~Fang, E.~Liang, and R.~Zhong, ``Hierarchical control for stochastic network traffic with reinforcement learning,'' \emph{Transportation Research Part B: Methodological}, vol. 167, pp. 196--216, 2023.

\bibitem{chow2021adaptive}
A.~H. Chow, Z.~Su, E.~Liang, and R.~Zhong, ``Adaptive signal control for bus service reliability with connected vehicle technology via reinforcement learning,'' \emph{Transportation Research Part C: Emerging Technologies}, vol. 129, p. 103264, 2021.

\bibitem{lighthill1955kinematic}
M.~J. Lighthill and G.~B. Whitham, ``On kinematic waves ii. {A} theory of traffic flow on long crowded roads,'' \emph{Proceedings of the Royal Society A}, vol. 229, no. 1178, pp. 317--345, 1955.

\bibitem{su2021adaptive}
Z.~Su, A.~H. Chow, and R.~Zhong, ``Adaptive network traffic control with an integrated model-based and data-driven approach and a decentralised solution method,'' \emph{Transportation Research Part C: Emerging Technologies}, vol. 128, p. 103154, 2021.

\bibitem{moridpour2010lane}
S.~Moridpour, M.~Sarvi, and G.~Rose, ``Lane changing models: {A} critical review,'' \emph{Transportation Letters}, vol.~2, no.~3, pp. 157--173, 2010.

\bibitem{TransCAD}
\BIBentryALTinterwordspacing
Caliper. (2023) {TransCAD}. \url{https://www.caliper.com/tcovu.htm}. Accessed on July 23, 2024. [Online]. Available: \url{https://www.caliper.com/tcovu.htm}
\BIBentrySTDinterwordspacing

\bibitem{EMME}
\BIBentryALTinterwordspacing
INRO. (2023) {EMME}. \url{https://www.inrosoftware.com/en/products/emme/}. Accessed on July 23, 2024. [Online]. Available: \url{https://www.inrosoftware.com/en/products/emme/}
\BIBentrySTDinterwordspacing

\bibitem{PTV-Visum}
\BIBentryALTinterwordspacing
P.~Group. (2023) Ptv-visum. \url{https://www.ptvgroup.com/zh-hant/products/ptv-visum}. Accessed on July 23, 2024. [Online]. Available: \url{https://www.ptvgroup.com/zh-hant/products/ptv-visum}
\BIBentrySTDinterwordspacing

\bibitem{DBLP:journals/tvcg/DengWLBZSXW22}
Z.~Deng, D.~Weng, Y.~Liang, J.~Bao, Y.~Zheng, T.~Schreck, M.~Xu, and Y.~Wu, ``Visual cascade analytics of large-scale spatiotemporal data,'' \emph{{IEEE Trans. Vis. Comput. Graph.}}, vol.~28, no.~6, pp. 2486--2499, 2022.

\bibitem{DBLP:journals/tvcg/LandesbergerBRA16}
T.~von Landesberger, F.~Brodkorb, P.~Roskosch, N.~V. Andrienko, G.~L. Andrienko, and A.~Kerren, ``{MobilityGraphs}: {V}isual analysis of mass mobility dynamics via spatio-temporal graphs and clustering,'' \emph{{IEEE Trans. Vis. Comput. Graph.}}, vol.~22, no.~1, pp. 11--20, 2016.

\bibitem{DBLP:journals/tvcg/WengZDMBZXW21}
D.~Weng, C.~Zheng, Z.~Deng, M.~Ma, J.~Bao, Y.~Zheng, M.~Xu, and Y.~Wu, ``Towards better bus networks: {A} visual analytics approach,'' \emph{{IEEE Trans. Vis. Comput. Graph.}}, vol.~27, no.~2, pp. 817--827, 2021.

\bibitem{DBLP:journals/tvcg/LorenzoSCBPN16}
G.~D. Lorenzo, M.~L. Sbodio, F.~Calabrese, M.~Berlingerio, F.~Pinelli, and R.~Nair, ``{AllAboard}: {V}isual exploration of cellphone mobility data to optimise public transport,'' \emph{{IEEE Trans. Vis. Comput. Graph.}}, vol.~22, no.~2, pp. 1036--1050, 2016.

\bibitem{DBLP:journals/jvis/DengWW23}
Z.~Deng, D.~Weng, and Y.~Wu, ``You are experienced: {I}nteractive tour planning with crowdsourcing tour data from web,'' \emph{J. Vis.}, vol.~26, no.~2, pp. 385--401, 2023.

\bibitem{DBLP:conf/visualization/LiuLTLMC20}
Q.~Q. Liu, Q.~Li, C.~F. Tang, H.~Lin, X.~Ma, and T.~Chen, ``A visual analytics approach to scheduling customized shuttle buses via perceiving passengers' travel demands,'' in \emph{Proceedings of {IEEE} Visualization Conference}, 2020, pp. 76--80.

\bibitem{DBLP:journals/tvcg/JinLPCTCK23}
S.~Jin, H.~Lee, C.~Park, H.~Chu, Y.~Tae, J.~Choo, and S.~Ko, ``A visual analytics system for improving attention-based traffic forecasting models,'' \emph{{IEEE} Trans. Vis. Comput. Graph.}, vol.~29, no.~1, pp. 1102--1112, 2023.

\bibitem{FSLens}
L.~Chen, H.~Wang, Y.~Ouyang, Y.~Zhou, N.~Wang, and Q.~Li, ``{FSLens}: {A} visual analytics approach to evaluating and optimizing the spatial layout of fire stations,'' \emph{{IEEE Trans. Vis. Comput. Graph.}}, vol.~30, no.~1, pp. 847--857, 2024.

\bibitem{DBLP:journals/tvcg/WengCDWCW19}
D.~Weng, R.~Chen, Z.~Deng, F.~Wu, J.~Chen, and Y.~Wu, ``{SRVis}: {T}owards better spatial integration in ranking visualization,'' \emph{{IEEE Trans. Vis. Comput. Graph.}}, vol.~25, no.~1, pp. 459--469, 2019.

\bibitem{simon1960new}
H.~A. Simon, ``The new science of management decision,'' 1960.

\bibitem{DecisionMaking}
E.~Oral, R.~Chawla, M.~Wijkstra, N.~Mahyar, and E.~Dimara, ``From information to choice: {A} critical inquiry into visualization tools for decision making,'' \emph{{IEEE Trans. Vis. Comput. Graph.}}, vol.~30, no.~1, pp. 359--369, 2024.

\bibitem{DBLP:journals/ivs/HanS23}
W.~Han and H.~Schulz, ``Providing visual analytics guidance through decision support,'' \emph{Information Visualization}, vol.~22, no.~2, pp. 140--165, 2023.

\bibitem{TrafficSurvey}
W.~Chen, F.~Guo, and F.-Y. Wang, ``A survey of traffic data visualization,'' \emph{{IEEE Trans. Intell. Transp. Syst.}}, vol.~16, no.~6, pp. 2970--2984, 2015.

\bibitem{BusSchedules}
C.~Palomo, Z.~Guo, C.~T. Silva, and J.~Freire, ``Visually exploring transportation schedules,'' \emph{IEEE Trans. Vis. Comput. Graph.}, vol.~22, no.~1, pp. 170--179, 2016.

\bibitem{ZengPublicTraffic}
W.~Zeng, C.-W. Fu, S.~M. Arisona, A.~Erath, and H.~Qu, ``Visualizing mobility of public transportation system,'' \emph{IEEE Trans. Vis. Comput. Graph.}, vol.~20, no.~12, pp. 1833--1842, 2014.

\bibitem{DBLP:journals/tvcg/SunLQW17}
G.~Sun, R.~Liang, H.~Qu, and Y.~Wu, ``Embedding spatio-temporal information into maps by route-zooming,'' \emph{{IEEE Trans. Vis. Comput. Graph.}}, vol.~23, no.~5, pp. 1506--1519, 2017.

\bibitem{CongestionVis}
C.~Lee, Y.~Kim, S.~Jin, D.~Kim, R.~Maciejewski, D.~Ebert, and S.~Ko, ``A visual analytics system for exploring, monitoring, and forecasting road traffic congestion,'' \emph{{IEEE Trans. Vis. Comput. Graph.}}, vol.~26, no.~11, pp. 3133--3146, 2020.

\bibitem{DBLP:journals/cgf/SchottlerYPB21}
S.~Sch{\"{o}}ttler, Y.~Yang, H.~Pfister, and B.~Bach, ``Visualizing and interacting with geospatial networks: {A} survey and design space,'' \emph{Comput. Graph. Forum}, vol.~40, no.~6, pp. 5--33, 2021.

\bibitem{DBLP:conf/iv/MeninCFCW21}
A.~Menin, R.~A. Cava, C.~M. D.~S. Freitas, O.~Corby, and M.~Winckler, ``Towards a visual approach for representing analytical provenance in exploration processes,'' in \emph{Proceedings of International Conference Information Visualisation}.\hskip 1em plus 0.5em minus 0.4em\relax {IEEE}, 2021, pp. 21--28.

\bibitem{DBLP:journals/cgf/XuOWSCW20}
K.~Xu, A.~Ottley, C.~Walchshofer, M.~Streit, R.~Chang, and J.~E. Wenskovitch, ``Survey on the analysis of user interactions and visualization provenance,'' \emph{Comput. Graph. Forum}, vol.~39, no.~3, pp. 757--783, 2020.

\bibitem{DBLP:journals/tvcg/RaganESC16}
E.~D. Ragan, A.~Endert, J.~Sanyal, and J.~Chen, ``Characterizing provenance in visualization and data analysis: An organizational framework of provenance types and purposes,'' \emph{{IEEE} Trans. Vis. Comput. Graph.}, vol.~22, no.~1, pp. 31--40, 2016.

\bibitem{DBLP:journals/tvcg/Nguyen0WWAF16}
P.~H. Nguyen, K.~Xu, A.~Wheat, B.~L.~W. Wong, S.~Attfield, and B.~Fields, ``{SensePath}: {Understanding} the sensemaking process through analytic provenance,'' \emph{{IEEE} Trans. Vis. Comput. Graph.}, vol.~22, no.~1, pp. 41--50, 2016.

\bibitem{DBLP:conf/uist/GrossmanMF10}
T.~Grossman, J.~Matejka, and G.~W. Fitzmaurice, ``{Chronicle}: {C}apture, exploration, and playback of document workflow histories,'' in \emph{Proceedings of the Annual {ACM} Symposium on User Interface Software and Technology}, 2010, pp. 143--152.

\bibitem{DBLP:journals/tvcg/BorkinYBMGSP13}
M.~A. Borkin, C.~S. Yeh, M.~Boyd, P.~Macko, K.~Z. Gajos, M.~I. Seltzer, and H.~Pfister, ``Evaluation of filesystem provenance visualization tools,'' \emph{{IEEE} Trans. Vis. Comput. Graph.}, vol.~19, no.~12, pp. 2476--2485, 2013.

\bibitem{DBLP:journals/tvcg/StitzGPZS19}
H.~Stitz, S.~Gratzl, H.~Piringer, T.~Zichner, and M.~Streit, ``{KnowledgePearls}: Provenance-based visualization retrieval,'' \emph{{IEEE} Trans. Vis. Comput. Graph.}, vol.~25, no.~1, pp. 120--130, 2019.

\bibitem{DBLP:journals/tvcg/YangSDTLT17}
X.~Yang, L.~Shi, M.~Daianu, H.~Tong, Q.~Liu, and P.~M. Thompson, ``Blockwise human brain network visual comparison using nodetrix representation,'' \emph{{IEEE} Trans. Vis. Comput. Graph.}, vol.~23, no.~1, pp. 181--190, 2017.

\bibitem{DBLP:journals/jvis/JinC0QXC21}
Z.~Jin, N.~Chen, Y.~Shi, W.~Qian, M.~Xu, and N.~Cao, ``{TrammelGraph}: {V}isual graph abstraction for comparison,'' \emph{J. Vis.}, vol.~24, no.~2, pp. 365--379, 2021.

\bibitem{DBLP:journals/tvcg/ShiHTTDJWT22}
L.~Shi, J.~Hu, Z.~Tan, J.~Tao, J.~Ding, Y.~Jin, Y.~Wu, and P.~M. Thompson, ``{MVNet}: {M}ulti-variate multi-view brain network comparison over uncertain data,'' \emph{{IEEE} Trans. Vis. Comput. Graph.}, vol.~28, no.~12, pp. 4640--4657, 2022.

\bibitem{DBLP:journals/tvcg/YoghourdjianDKM18}
V.~Yoghourdjian, T.~Dwyer, K.~Klein, K.~Marriott, and M.~Wybrow, ``{Graph Thumbnails}: {I}dentifying and comparing multiple graphs at a glance,'' \emph{{IEEE} Trans. Vis. Comput. Graph.}, vol.~24, no.~12, pp. 3081--3095, 2018.

\bibitem{DBLP:conf/vinci/BurchW14}
M.~Burch and D.~Weiskopf, ``A flip-book of edge-splatted small multiples for visualizing dynamic graphs,'' in \emph{Proceedings of the International Symposium on Visual Information Communication and Interaction}.\hskip 1em plus 0.5em minus 0.4em\relax {ACM}, 2014, p.~29.

\bibitem{DBLP:journals/cgf/BachRDMFG15}
B.~Bach, N.~H. Riche, T.~Dwyer, T.~M. Madhyastha, J.~Fekete, and T.~J. Grabowski, ``{Small MultiPiles}: {P}iling time to explore temporal patterns in dynamic networks,'' \emph{Comput. Graph. Forum}, vol.~34, no.~3, pp. 31--40, 2015.

\bibitem{DBLP:journals/cgf/CornelKSHBVW16}
D.~Cornel, A.~Konev, B.~Sadransky, Z.~Horv{\'{a}}th, A.~Brambilla, I.~Viola, and J.~Waser, ``Composite flow maps,'' \emph{Computer Graphics Forum}, vol.~35, no.~3, pp. 461--470, 2016.

\bibitem{mcnally2007four}
M.~G. McNally, ``The four-step model,'' in \emph{Handbook of Transport Modelling}, 2007, vol.~1, pp. 35--53.

\bibitem{lu2024two}
Q.-L. Lu, M.~Qurashi, and C.~Antoniou, ``A two-stage stochastic programming approach for dynamic od estimation using lbsn data,'' \emph{Transportation Research Part C: Emerging Technologies}, vol. 158, p. 104460, 2024.

\bibitem{wang2013global}
S.~Wang, Q.~Meng, and H.~Yang, ``Global optimization methods for the discrete network design problem,'' \emph{Transportation Research Part B: Methodological}, vol.~50, pp. 42--60, 2013.

\bibitem{spiess1990conical}
H.~Spiess, ``Conical volume-delay functions,'' \emph{Transportation Science}, vol.~24, no.~2, pp. 153--158, 1990.

\bibitem{BRP}
{Bureau of Public Roads}, ``Traffic assignment manual,'' \emph{U.S. Dept. of Commerce, Ubran Planning Division, Washington, DC, USA, 1964}.

\bibitem{sheffi1982algorithm}
Y.~Sheffi and W.~B. Powell, ``An algorithm for the equilibrium assignment problem with random link times,'' \emph{Networks}, vol.~12, no.~2, pp. 191--207, 1982.

\bibitem{liu2009method}
H.~X. Liu, X.~He, and B.~He, ``Method of successive weighted averages (mswa) and self-regulated averaging schemes for solving stochastic user equilibrium problem,'' \emph{Networks and Spatial Economics}, vol.~9, pp. 485--503, 2009.

\bibitem{DBLP:journals/tvcg/SedlmairMM12}
M.~Sedlmair, M.~D. Meyer, and T.~Munzner, ``Design study methodology: {R}eflections from the trenches and the stacks,'' vol.~18, no.~12, pp. 2431--2440, 2012.

\bibitem{DBLP:journals/tvcg/PiYSJ21}
M.~Pi, H.~Yeon, H.~Son, and Y.~Jang, ``Visual cause analytics for traffic congestion,'' \emph{{IEEE Trans. Vis. Comput. Graph.}}, vol.~27, no.~3, pp. 2186--2201, 2021.

\bibitem{cibulski2022supporting}
L.~Cibulski, E.~Dimara, S.~Hermawati, and J.~Kohlhammer, ``Supporting domain characterization in visualization design studies with the critical decision method,'' in \emph{Proceedings of IEEE Workshop on Visualization Guidelines in Research, Design, and Education}, 2022, pp. 8--15.

\bibitem{DBLP:journals/tvcg/YangDGM17}
Y.~Yang, T.~Dwyer, S.~Goodwin, and K.~Marriott, ``Many-to-many geographically-embedded flow visualisation: {A}n evaluation,'' \emph{{IEEE Trans. Vis. Comput. Graph.}}, vol.~23, no.~1, pp. 411--420, 2017.

\bibitem{DBLP:journals/cgf/ZengSJT19}
W.~Zeng, Q.~Shen, Y.~Jiang, and A.~C. Telea, ``Route-aware edge bundling for visualizing origin-destination trails in urban traffic,'' \emph{Comput. Graph. Forum}, vol.~38, no.~3, pp. 581--593, 2019.

\bibitem{DBLP:journals/ivs/GleicherAWJHR11}
M.~Gleicher, D.~Albers, R.~Walker, I.~Jusufi, C.~D. Hansen, and J.~C. Roberts, ``Visual comparison for information visualization,'' \emph{Information Visualization}, vol.~10, no.~4, pp. 289--309, 2011.

\bibitem{braess1968paradoxon}
D.~Braess, ``{\"U}ber ein paradoxon aus der verkehrsplanung,'' \emph{Unternehmensforschung}, vol.~12, pp. 258--268, 1968.

\bibitem{steinberg1983prevalence}
R.~Steinberg and W.~I. Zangwill, ``The prevalence of braess' paradox,'' \emph{Transportation Science}, vol.~17, no.~3, pp. 301--318, 1983.

\bibitem{leblanc1975efficient}
L.~J. LeBlanc, E.~K. Morlok, and W.~P. Pierskalla, ``An efficient approach to solving the road network equilibrium traffic assignment problem,'' \emph{Transportation research}, vol.~9, no.~5, pp. 309--318, 1975.

\bibitem{leblanc1975algorithm}
L.~J. Leblanc, ``An algorithm for the discrete network design problem,'' \emph{Transportation Science}, vol.~9, no.~3, pp. 183--199, 1975.

\bibitem{di2020reversible}
Z.~Di and L.~Yang, ``Reversible lane network design for maximizing the coupling measure between demand structure and network structure,'' \emph{Transportation Research Part E: Logistics and Transportation Review}, vol. 141, p. 102021, 2020.

\bibitem{yin2022simulation}
R.~Yin, X.~Liu, N.~Zheng, and Z.~Liu, ``Simulation-based analysis of second-best multimodal network capacity,'' \emph{Transportation Research Part C: Emerging Technologies}, vol. 145, p. 103925, 2022.

\bibitem{DBLP:journals/cgf/PerinVSSWC18}
C.~Perin, R.~Vuillemot, C.~D. Stolper, J.~T. Stasko, J.~Wood, and S.~Carpendale, ``State of the art of sports data visualization,'' \emph{Comput. Graph. Forum}, vol.~37, no.~3, pp. 663--686, 2018.

\bibitem{DBLP:journals/tmm/WuCGTK16}
Y.~Wu, N.~Cao, D.~Gotz, Y.~Tan, and D.~A. Keim, ``A survey on visual analytics of social media data,'' \emph{{IEEE} Trans. Multim.}, vol.~18, no.~11, pp. 2135--2148, 2016.

\bibitem{JAFARI2017270}
E.~Jafari, V.~Pandey, and S.~D. Boyles, ``A decomposition approach to the static traffic assignment problem,'' \emph{Transportation Research Part B: Methodological}, vol. 105, pp. 270--296, 2017.

\bibitem{raadsen2020aggregation}
M.~P. Raadsen, M.~C. Bliemer, and M.~G. Bell, ``Aggregation, disaggregation and decomposition methods in traffic assignment: historical perspectives and new trends,'' \emph{Transportation research part B: methodological}, vol. 139, pp. 199--223, 2020.

\bibitem{yahia2018network}
C.~N. Yahia, V.~Pandey, and S.~D. Boyles, ``Network partitioning algorithms for solving the traffic assignment problem using a decomposition approach,'' \emph{Transportation Research Record}, vol. 2672, no.~48, pp. 116--126, 2018.

\end{thebibliography}


 


\begin{IEEEbiography}[{\includegraphics[width=1in,height=1.25in,clip,keepaspectratio]{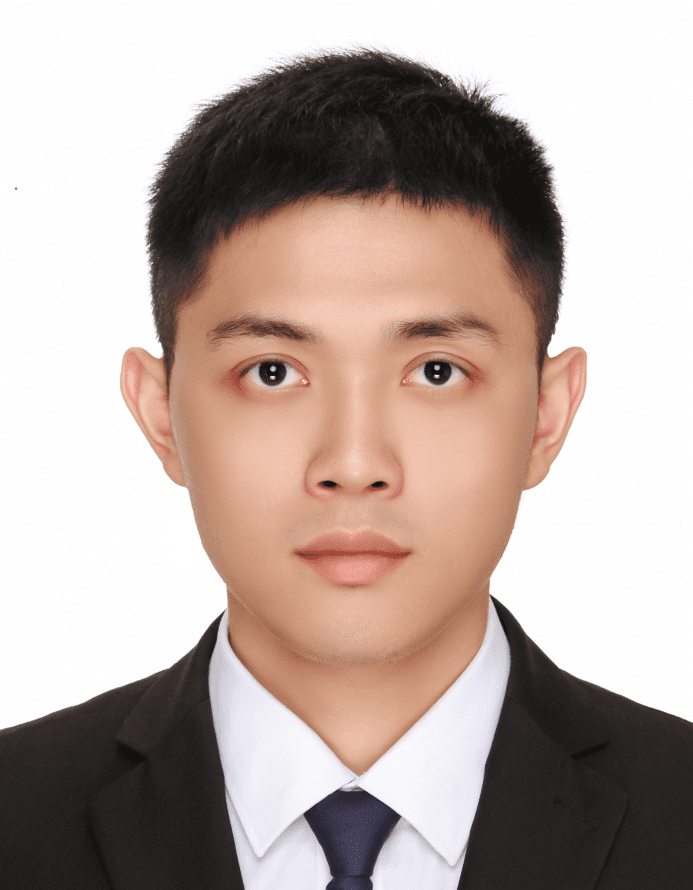}}]{Zikun Deng} is a tenure-track associate professor at School of Software Engineering, South China University of Technology. He received his Ph.D. degree in Computer Science from the State Key Lab of CAD\&CG, Zhejiang University in 2023. His research interests mainly include visual analytics, visualization, data mining, and their application in smart city, industry 4.0, and digital twins. He has published more than 10 papers in IEEE TVCG. For more information, please visit https://zkdeng.org.
\end{IEEEbiography}

\begin{IEEEbiography}[{\includegraphics[width=1in,height=1.25in,clip,keepaspectratio]{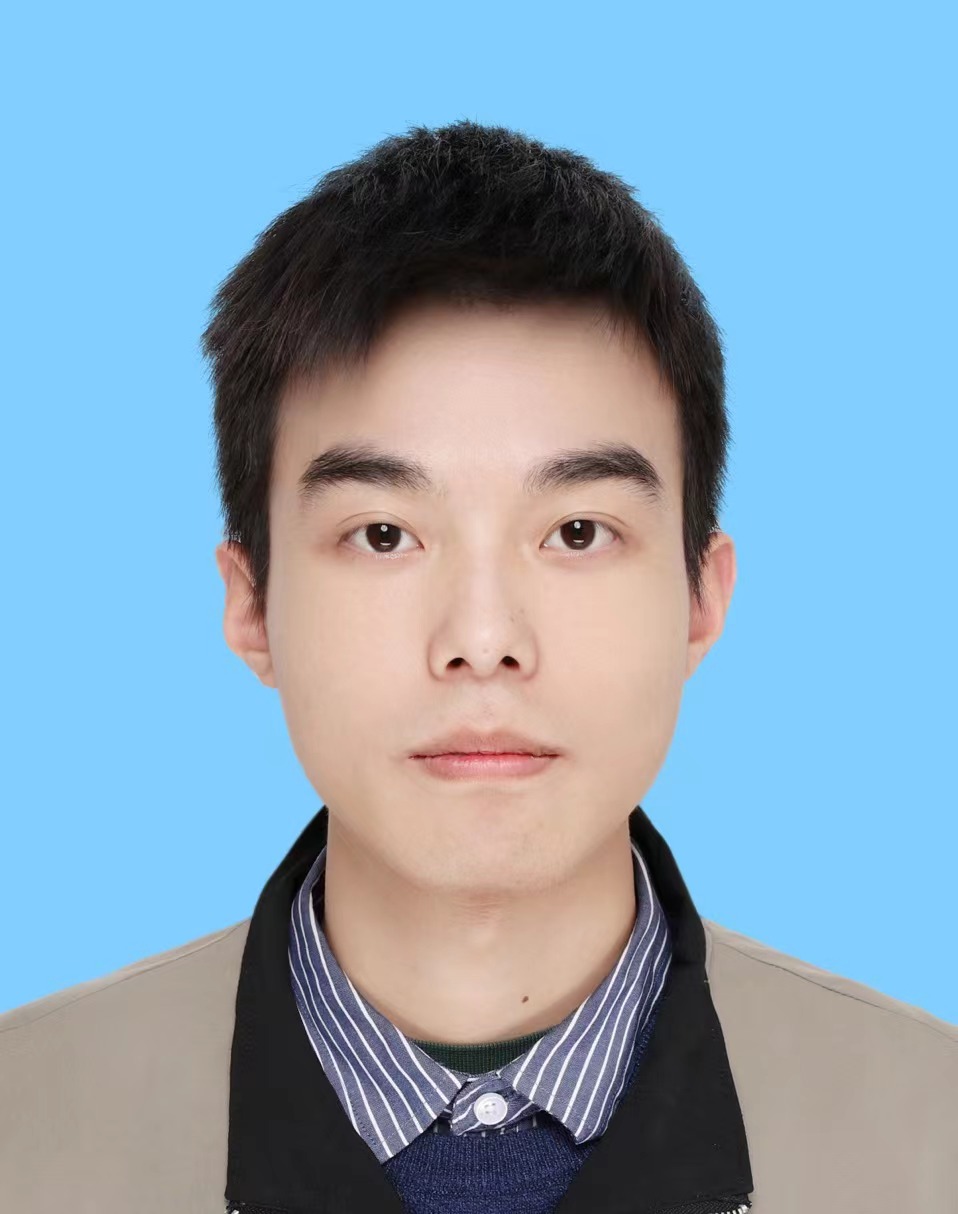}}]{Yuanbang Liu} is an undergraduate student at the School of Software Engineering, South China University of Technology, and his research interests are data visualization and visual analytics.
\end{IEEEbiography}

\begin{IEEEbiography}[{\includegraphics[width=1in,height=1.25in,clip,keepaspectratio]{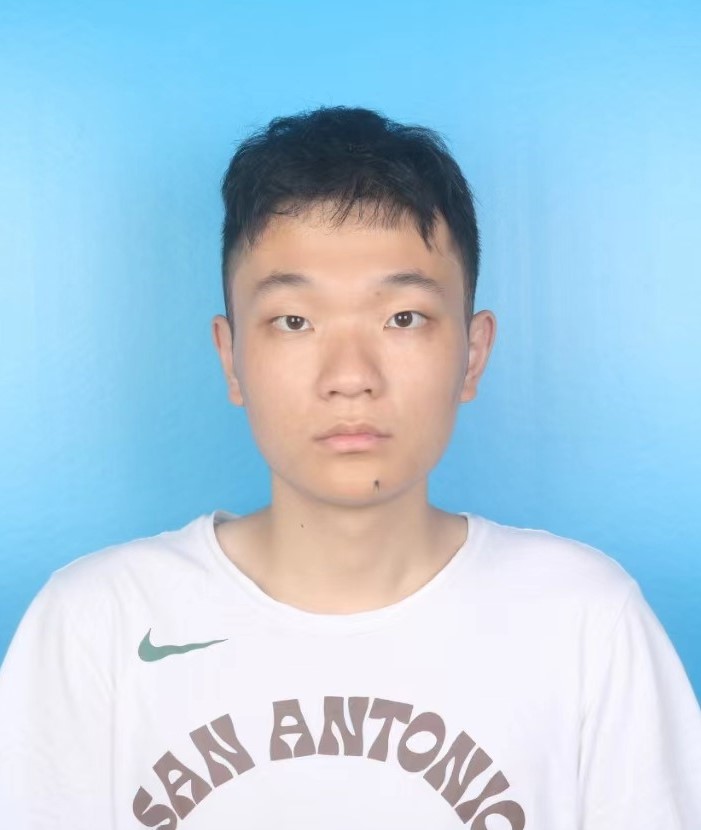}}]{Mingrui Zhu} is an undergraduate student at the School of Software Engineering, South China University of Technology, and his research interests are data visualization and visual analytics.
\end{IEEEbiography}

\begin{IEEEbiography}[{\includegraphics[width=1in,height=1.25in,clip,keepaspectratio]{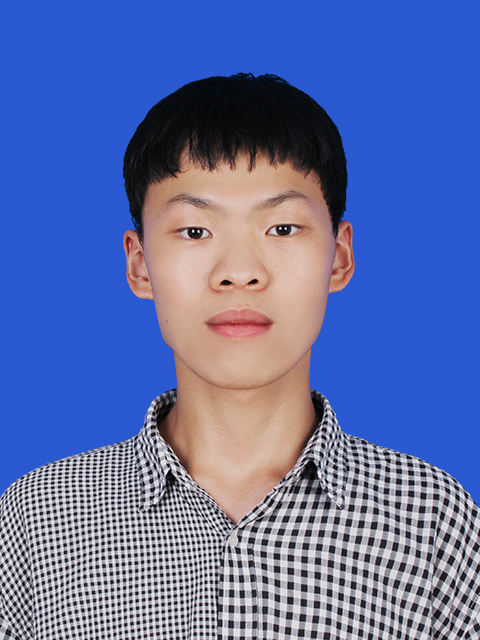}}]{Da Xiang} is an undergraduate student at the School of Software Engineering, South China University of Technology, and his research interests are data visualization and visual analytics.
\end{IEEEbiography}

\begin{IEEEbiography}[{\includegraphics[width=1in,height=1.25in,clip,keepaspectratio]{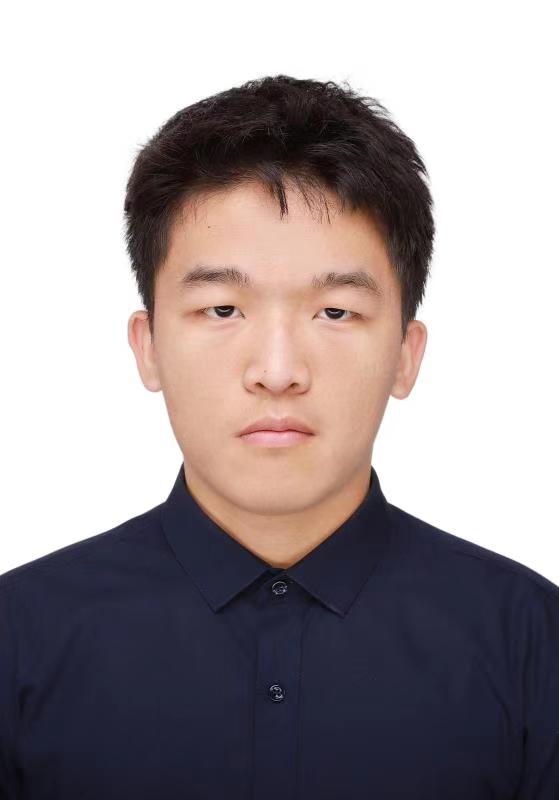}}]{Haiyue Yu} is an undergraduate student at the School of Software Engineering, South China University of Technology, and his research interests are data visualization and visual analytics.
\end{IEEEbiography}

\begin{IEEEbiography}[{\includegraphics[width=1in,height=1.25in,clip,keepaspectratio]{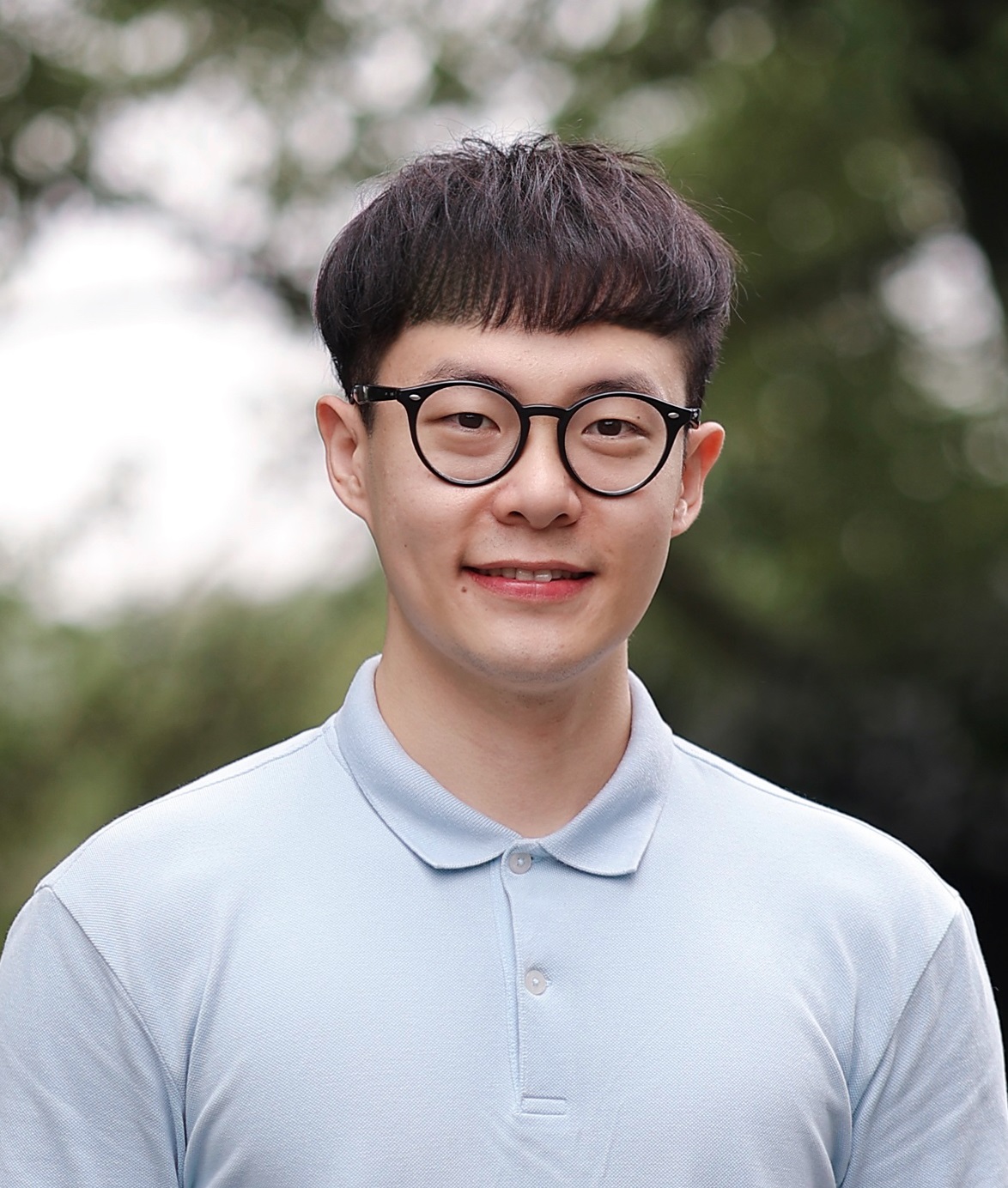}}]{Zicheng Su} is a distinguished researcher at Tongji University. His research interests include traffic signal control, traffic flow modeling, optimal control and reinforcement learning. Dr. Su has authored/co-authored more than ten peer-reviewed papers published in top transportation journals and conferences including Transportation Research Part B, Transportation Research Part C, ISTTT and AAAI. He has won the Outstanding Student Paper Award of Hong Kong Society for Transportation Studies (HKSTS) in 2021.
\end{IEEEbiography}

\begin{IEEEbiography}[{\includegraphics[width=1in,height=1.25in,clip,keepaspectratio]{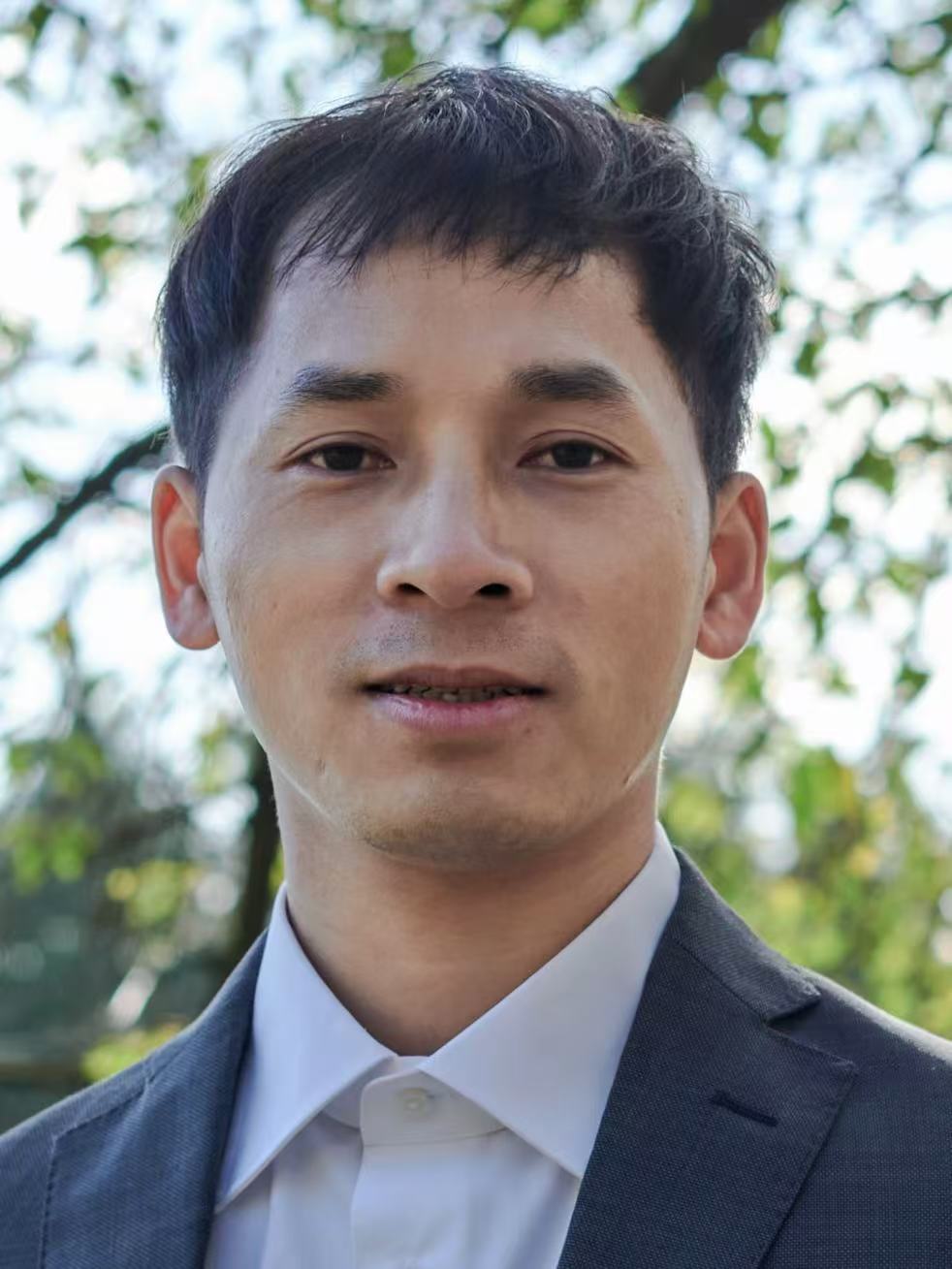}}]{Qinglong Lu} received the bachelor’s degree in traffic engineering from Sun Yat-sen University, Guangzhou, China, in 2018, and the master’s degree in transportation systems and the Ph.D. degree from the Technical University of Munich, Munich, Germany, in 2020 and 2024, respectively. He is a Postdoctoral Fellow with the National University of Singapore, Singapore. His research focuses on mobility pattern analysis, transportation system resilience evaluation, simulation-based optimization.
\end{IEEEbiography}

\begin{IEEEbiography}[{\includegraphics[width=1in,height=1.25in,clip,keepaspectratio]{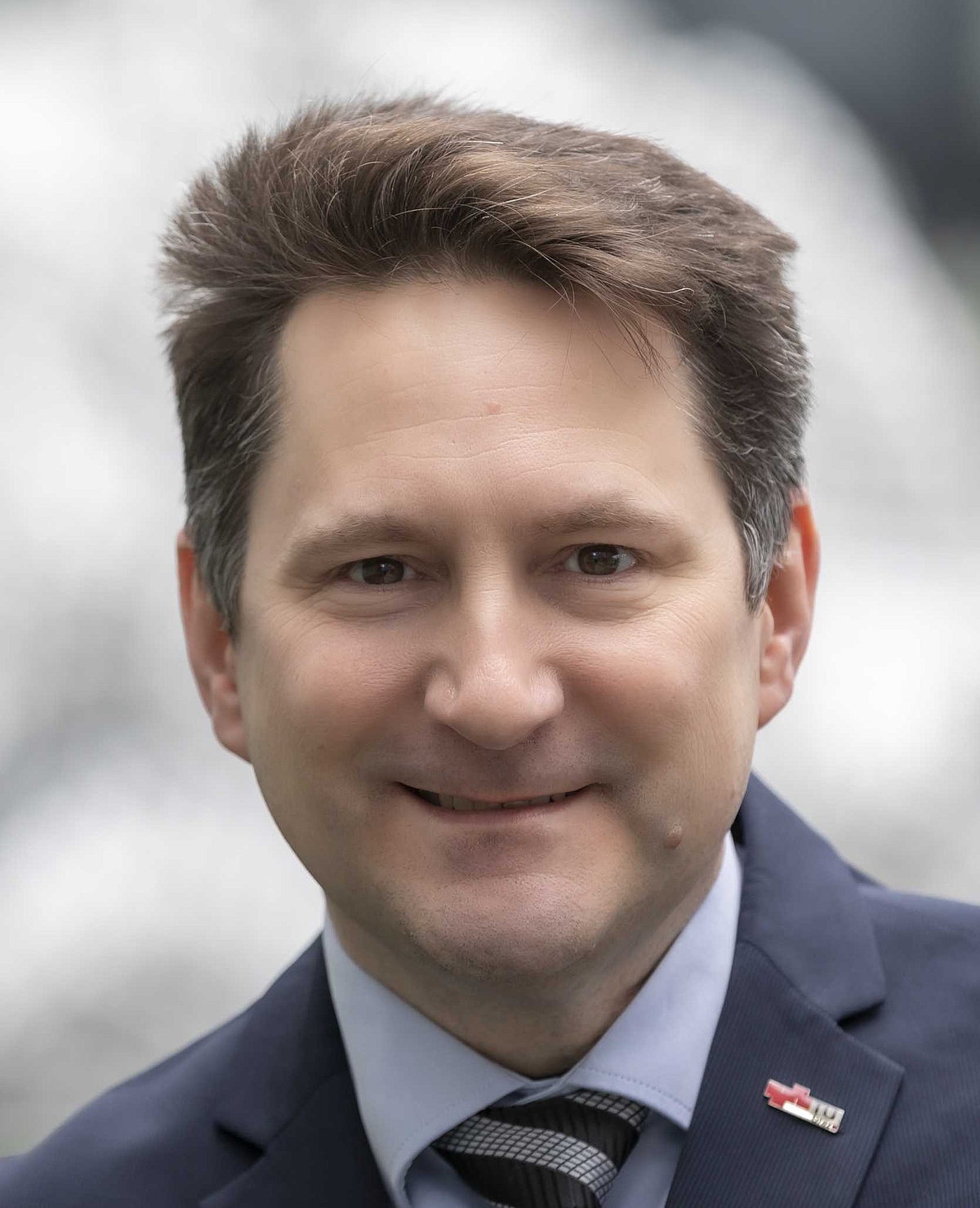}}]{Tobias Schreck} received the PhD degree in computer science in 2006 from the University of Konstanz. He is currently a professor and head of the Institute of Computer Graphics and Knowledge Visualization at the Faculty for Computer Science and Biomedical Engineering of Graz University of Technology. His main research interests are in Visual Data Analysis and in applied 3D Object Retrieval.
\end{IEEEbiography}

\begin{IEEEbiography}[{\includegraphics[width=1in,height=1.25in,clip,keepaspectratio]{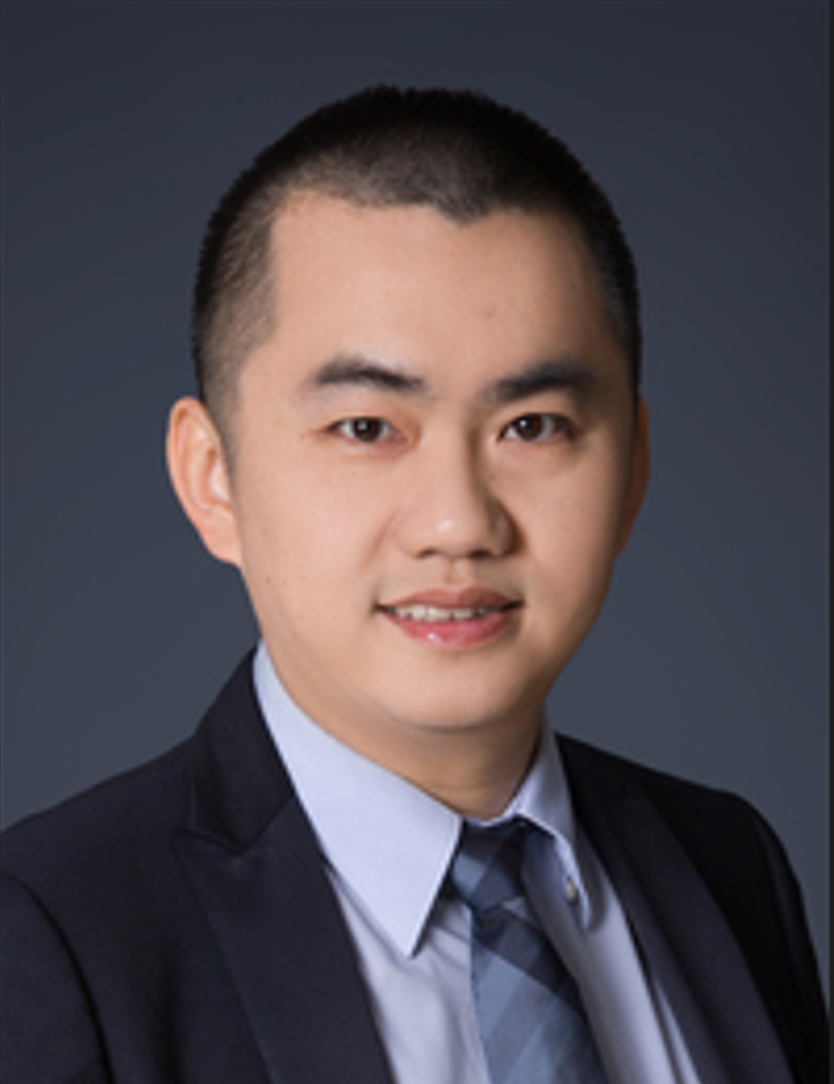}}]{Yi Cai} is the Dean of School of Software Engineering in South China University of Technology, the Director of The China Ministry of Education Key Laboratory of Big Data and Robotic Intelligence. He had received his PhD degree in the Chinese University of Hong Kong, and work as postdoctoral fellow in City University of Hong Kong. He has published more than 170 high quality papers in top journals and conferences such as IEEE TVCG, TKDE, TMM, AAAI, ACL, and ACM MM. He also has published 2 academic monographs and acts as the Chairman and Program Committee members of more than 20 prestigious international academic conferences.
\end{IEEEbiography}



\vfill

\end{document}